\documentclass[twocolumn]{aastex631}
\usepackage{amsmath}
\usepackage{hyperref}
\begin{document}

\title{Primordial Black Hole Triggered Type Ia Supernovae II: Comparison with Supernova Remnants and Galactic Chemical Evolution}

\shortauthors{Leung, Walther, Kusenko, Nomoto, and Suzuki}
\shorttitle{PBH Triggered Type Ia Supernovae II: SNR and GCE}

\author[0000-0002-4972-3803]{Shing-Chi Leung}

\affiliation{Department of Physics, SUNY Polytechnic Institute, 100 Seymour Road, Utica, NY 13502, USA}

\affiliation{Kavli Institute for the Physics and 
Mathematics of the Universe (WPI), The University 
of Tokyo Institutes for Advanced Study, The 
University of Tokyo, Kashiwa, Chiba 277-8583, Japan}

\author[0009-0006-8467-2163]{Seth Walther}

\affiliation{Department of Electrical and Computer Engineering, SUNY Polytechnic Institute, 100 Seymour Road, Utica, NY 13502, USA}
\affiliation{Department of Mathematics, SUNY Polytechnic Institute, 100 Seymour Road, Utica, NY 13502, USA}
\affiliation{Department of Physics, SUNY Polytechnic Institute, 100 Seymour Road, Utica, NY 13502, USA}

\author[0000-0002-8619-1260]{Alexander Kusenko}

\affiliation{Department of Physics and Astronomy, University of California, Los Angeles California, 90095-1547, USA}

\affiliation{Kavli Institute for the Physics and 
Mathematics of the Universe (WPI), The University 
of Tokyo Institutes for Advanced Study, The 
University of Tokyo, Kashiwa, Chiba 277-8583, Japan}

\author[0000-0001-9553-0685]{Ken'ichi Nomoto}

\affiliation{Kavli Institute for the Physics and 
Mathematics of the Universe (WPI), The University 
of Tokyo Institutes for Advanced Study, The 
University of Tokyo, Kashiwa, Chiba 277-8583, Japan}

\author{Tomoharu Suzuki}
\affiliation{School of General Education, Chubu University, 1200 Matsumoto-cho, Kasugai, Aichi 487-8501, Japan}

\correspondingauthor{Shing-Chi Leung}
\email{leungs@sunypoly.edu}

\date{\today}

\submitjournal{ApJ}
\received{Feb 5 2026}
\revised{May 1 2026}
\accepted{May 12 2026}
published{Jun 15 2026}

\begin{abstract}

The asteroid-mass class of Primordial Black Holes (PBHs) is one of the candidates for the dark matter in the universe. With a mass between $4 \times 10^{-17} < M_{\rm PBH} < 4 \times 10^{-12}~M_{\odot}$, they could be the major component of dark matter in the cosmic mass budget. The infall of these PBH into a white dwarf could be one triggering mechanism of Type Ia supernovae (SNe Ia). In [Leung et al, ApJ 991, 11 (2025)] (Paper I), we studied the ignition, explosion dynamics, radiative transfer, and post-explosion nucleosynthesis of the PBH-triggered SNe Ia. The diversity of the explosion models can reconcile with the empirical Phillips relation. In this work, we developed the PBH-triggered SN Ia models in various metallicity. We show that models from this channel can explain some recently observed SN Ia light curves and supernova remnants. We further investigate how these supernovae could affect the chemical evolution on the galactic scale by adding the new SN Ia models as a new chemical source. We examine how the observed chemical trends of stars can lead to constraints on the fraction of this explosion channel relative to the canonical binary star channel. Our models suggest that the PBH can be one major SN Ia channel in the early universe. We also include a comparative study to extract the effects of PBH-triggered SN Ia parameters on the actual chemical trends in the galactic chemical evolution model.

\end{abstract}



\newcommand{\red}[1]{\textcolor{red}{#1}}



\section{Introduction}

Primordial black holes~\citep{Zeldovich:1967,Hawking:1971ei,Carr:1974nx}, which could have formed in the early universe before the formation of stars and large-scale structure, present an appealing candidate for dark matter~\citep{Green:2020jor,Carr:2020xqk}.  In the mass range $4 \times 10^{-17} \lesssim M_{\rm PBH} \lesssim 4 \times 10^{-12}~M_{\odot}$, PBHs could account for all dark matter.  While this mass range is difficult to explore, there are already some  observational hints for small PBHs in microlensing, r-process data, and other astrophysical data~\citep{Bean:2002kx,Kawasaki:2012kn,Belotsky:2014kca,Niikura:2017zjd,Fuller:2017uyd,Carr:2019kxo,Takhistov:2020vxs,Coogan:2020tuf,Sugiyama:2021xqg,Lu:2023xoi,Carr:2023tpt,Flores:2023lll}.  

The astroid mass window, in which there are currently no constrains, is inaccessible to microlensing observations in light or in gamma rays because of the combination of finite size source effects and wave effects~\citep{Katz:2018zrn,Sugiyama:2019dgt,Smyth:2019whb,Profumo:2025uwc}.  Precisions monitoring of ephemerides could help discover a PBH in this mass range~\citep{Tran:2023jci}.  

A promising approach is to use a combination of compact stars, such as neutron stars~\citep{Fuller:2017uyd} and white dwarfs~\citep{Graham:2015apa}, to discover dark-matter PBHs by their effects on these compact stars.  When a PBH is captured by a neutron star, it consumes the neutron star from the inside and produces a kilonova-like event accompanied by a fast radio burst (FRB), but not accompanied by a gravitational-waves (GW) signature of merging compact stars~\citep{Fuller:2017uyd}.  Therefore, an observation of a kilonova accompanied by an FRB but not triggered by a GW detector would be a signature of PBH dark matter~\citep{Fuller:2017uyd}.  
The remnants of neutron stars consumed by PBHs are 1-2 $M_\odot$ black holes, which can be detected by the GW detectors~\citep{Fuller:2017uyd,Takhistov:2017bpt,Takhistov:2020vxs}, and which may have already been observed in the galactic center as G-objects~\citep{Flores:2023lll}.

When a PBH passes through a white dwarf (WD), it can generate enough heat to initiate a supernova explosion~\citep{Graham:2015apa,Leung2025DMPBH1} with unique spectral features, including a high yield of $^{56}$Ni.  To discover dark-matter PBHs or to set constraints, we need to identify the observational signatures of supernovae triggered by PBHs and to compare them with observational data. 

\subsection{Constraining SN Ia from Observations}

Type Ia supernovae (SNe Ia) are the explosions of C+O white dwarfs (WDs). On one hand, the Phillips relation \citep{Phillips1993AbsMag} casts SNe Ia naturally as a robust distance indicator \citep{Riess1995SNIaHubble, Riess1996DistanceIndicator}, which later fosters the discovery of dark energy \citep{Perlmutter1999DarkEnergy}. On the other hand, SNe Ia display vast diversity in terms of explosion strength and chemical and spectral variations \citep{Taubenberger2017HBSNReview}, which support multiple explosion scenarios. Many explosion mechanisms are thus proposed in the literature to span the possibilities of igniting the WDs. In general, it requires the thermonuclear runaways to exist in the form of deflagration \citep{Nomoto1976CDef} or detonation \citep{Arnett1969CDet}. These models are later developed into more sophisticated models including the turbulent deflagration model \citep{Niemeyer1995TurbFlame,Reinecke1999SNIa,Roepke2007CDefSNIa,Leung2020Iax}, deflagration-detonation transition model \citep{Khokhlov1991DDT, Khokhlov1997DDT, Gamezo2005DDT, Seitenzahl2013DDT, Leung2018Chand}, double detonation model \citep{Nomoto1982W7II,Livne1991DD,Woosley1994DD,Woosley2011DD,Sim2012DD,Shen2018DDDDDD, Leung2020SubChand}, gravitationally confined detonation model\footnote{Similar models are also denoted as pulsational reversed detonation model. See, e.g., \cite{Bravo2009GCD1}.} \citep{Bravo2009GCD, Jordan2008GCE, Jordan2012GCD}, violent merger \citep{Pakmor2017VioMerg}, and direct collision \citep{Kushnir2013Collision, Dong2015Collision}. These models span the binary interaction possibilities in how Chandrasekhar mass (Ch-mass) and sub-Chandrasekhar mass (subCh-mass) WDs explode in both single and double degenerate scenarios. 

Observations have supported or disputed individual mechanisms through the light curve or spectral fitting, and by the chemical abundance pattern \citep[see reviews, e.g., ][]{Hillebrandt2000SNIaReview,Hillebrandt2011SNIaReview, Nomoto2017HBSN-SNIa,Leung2023SNIaReview}. The detailed observation of stars, supernova light curves and spectra, supernova remnants (SNR), and the global trends of chemical elements in the Galaxy are common testbeds for these SN Ia models. 

The SNRs in the Milky Way can validate individual SN models. The X-ray spectra emitted by the hot gas in the ejecta reveal the Si-group elements (S, Ca), and Fe-group elements (Cr, Mn, Fe, Ni) that can be compared with the nucleosynthesis of the explosion models. Some well-observed SNRs which are SN Ia candidates include the Tycho \citep{Park2013Tycho}, Kepler \citep{Yamaguchi2014Kepler}, 3C 397 \citep{Yamaguchi20153C397,Ohshiro20213C397}, and Sagittarius East A \citep{Zhou2021SagA}. The comparison with individual models indicates the WD progenitors are Ch-mass or subCh-mass, and the initial metallicity. The G306.3-0.9 \citep{Wang2022G306.3-0.9} is one interesting example featuring the high Ca content. 
. 
Some nearby SNe Ia, e.g., SN 2011fe \citep{Nugent2011SN2011fe}, 2012cg \citep{Silverman2013SN2012cg}, 2014J \citep{Goobar2014SN2014J} and 2015F \citep{Im2015SN2015F,Cartier2017SN2015F} are extensively observed with multiple bandwidths. The light curves, including the peak luminosity, shape, and the late-time evolution, provide direct constraints on certain isotope pairs, e.g., $^{57}$Ni/$^{56}$Ni \citep{Churazov2014SN2014JLateLC, Graur2016SN2012cgLateLC, Graur2018SN2015F}, which could reveal the identity of their progenitors \citep{Mori2018SNIaNucleo, Tiwari2022SNIaLC}. Other observables, such as the H-line from companion stars \citep{Lunqvist2015SN2014JNoComp}, X-ray emission \citep{Margutti2014SN2014JNoXRay}, or archival image \citep{Li2011SN2011feNoComp}, could also leave constraints on the evolutionary path before the explosion (e.g., whether the SN Ia comes from a single degenerate or a double degenerate system). 

The Milky Way stars are another robust test place because stars of various metallicity are abundantly measured by star surveys \cite[e.g., APOGEE-2 in the SDSS IV, ][]{Majewski2017Apogee2}. How individual elements change as a function of metallicity reveals how families of supernovae behave collectively. This information provides us constraints on the relative importance of Ch-mass and subCh-mass WD in the SN Ia family, together with the delayed time \citep[see e.g., ][]{Ishigaki2021GaiaStar, Kobayashi2020GCE}. 

On an even larger scale, the hot gas in Galactic Clusters (e.g, the Perseus Cluster) plays a similar role. Billions of supernovae across the cosmic age explode and eject the synthesized heavy elements, which flow through galactic outflow to form the intragalactic medium. Different from individual galaxies, galactic clusters average out the effects of the local environment, and allow us to focus on the typical behaviour of the supernovae. In \cite{Simionescu2019Perseus}, the chemical abundance patterns in the Perseus Cluster measured by the Hitomi (Astro-H) telescope are extensively used to constrain the overall SN Ia rate and the Ch-mass fraction among SNe Ia.

\subsection{Motivation}

How DM affects SN Ia explosions can cast light on the nature of DM properties. The gravity of DM particles allows the WDs to explode beyond the conventional context. In \cite{Leung2015DMSNIa}, we showed that the DM can lower the Chandrasekhar mass of a WD even when the electron matter becomes highly degenerate. The lower mass progenitors lead to sub-luminous SNe Ia, which could match the light curves of some SNe Iax. In \cite{Chan2021DMSNIa}, we further show that the admixture of DM could explain some unusual low-luminosity slowly-evolving SNe Ia, in the parameter space orthogonal to the Phillips relation. In \cite{Leung2025DMPBH1} (Paper I), we further study the SN Ia explosions triggered by DM as accreting PBH. The PBH can create the hot spots and thus ignite the WD without binary interaction. We show that the model sequences of PBH-triggered SNe Ia can reproduce the Phillips relation.

Recent microlensing surveys such as MACHO project \citep[Massive Compact Halo Object, ][]{Alcock2000MACHO, Alcock2001MACHO}, EROS-2 \citep[Exp\/erience de Recherche d’Objets Sombres, ][]{Tisserand2007EROS2}, and the OGLE-5 \citep[Optical Gravitational Lensing Experiment, ][]{Niikura2019OGLE5} have significantly reduced the fraction of stellar-mass PBHs as a main component of cold dark matter (CDM). For example, only one microlensing event is observed by the Hyper Suprime-Cam \citep[HSC, ][]{Niikura2019HSC}. Meanwhile, PBH in the asteroid mass scale is not excluded. On top of that, a later reanalysis of the MACHO and EROS-2 results with an updated efficacy function shows that the PBHs stay important in the sub-solar mass scale between $10^{-4} - 10^2~M_{\odot}$ \citep{Garcia-Bellido2024PBH}. 

In \cite{MonteroCamacho2019DMSNIa}, the response of the WD due to the accretion of the astroid-mass PBH is studied assuming a 1D simulation with cylindrical symmetry. They showed that the studied mass scale is sufficient to trigger the explosion directly. With the potentially higher fraction of DM component as PBH, the PBH could be an alternative SN Ia explosion channel for a single WD in the early universe, when the overall DM density is much higher than nowadays. 
It leads to the question of how the theoretical possibility of PBH-triggered SNe Ia plays a role in the SN Ia family, as an explanation for the observed SN Ia light curves, and for the galactic chemical enrichment, especially in the early universe. On top of that, given that a single WD can explode directly as SNe Ia in the early universe, they tend to have a lower metallicity. This further questions how the nucleosynthesis for these SNe Ia depends on the WD metallicity and how it affects the metal enrichment process.  

In this article, we will first present the numerical methods in Section \ref{sec:methods}. Then in Section \ref{sec:results}, we present the metallicity-dependence of the SN Ia nucleosynthesis yields and compare the results with some well-observed supernova light curves and isotope mass fraction from SNRs. 
We apply our supernova nucleosynthesis yields to the Galactic Chemical Evolution (GCE) code in Section \ref{sec:GCE} and study the SN Ia models on the elemental trends. We study how some PBH-triggered SN Ia parameter affects individual elemental variation at different cosmic ages. 
Finally, in Section \ref{sec:discussion}, we discuss the caveat of our models, the implied supernova history, and then we give our conclusion. 

\section{Methods}
\label{sec:methods}

In Paper I, we reported the hydrodynamics simulations on the explosion phase of PBH-triggered SN Ia. In that work, we assume the PBH enters the WD by free-falling. We use the expected astroid mass scale of PBH being $\sim 10^{-12}~M_{\odot}$. During the passage of the PBH in the WD, the tidal interaction of the PBH and the accretion heating works as an external heat source. For moderate to extremely degenerate matter in a WD, the compressional heating could be sufficient to trigger self-sustaining C-burning, which quickly develops into a thermonuclear runaway. In MC19, they used one-dimensional cylindrical models and showed that a typical WD with a central density $> \sim 10^7$ g cm$^{-3}$ could be triggered into explosion. Furthermore, they predicted for a given WD mass, the impact parameter (enclosed mass) where the PBH can robustly trigger the SN Ia explosion. They have considered two extremes: whether the Kelvin-Helmholtz (KH) instabilities are important or not during the ignition phase. If KH instabilities are important, heat is efficiently removed by the outflow, which delays the ignition. Then the explosion will only happen at a higher mass WD regardless of the impact parameter. Otherwise, the WD is capable of being exploded at a lower mass.

In Paper I, we expanded our multi-dimensional compressible hydrodynamics code \citep{Leung2015SNCode}, which is primarily designed for modeling thermonuclear explosion \citep[e.g.,][]{Leung2018Chand, Leung2020SubChand, Leung2020Iax}, but the modular design of the input physics allows us to add additional physics \citep[see e.g., ][for applications in other types of supernovae.]{Leung2020ECSN,Leung2020SNOutburst,Leung2024Jet2}. Due to the rapid passage and the small mass of the PBH, the PBH is dynamically irrelevant to the later explosion. We use the results in MC19 and assume that the PBH has already induced the first flame at the given mass coordinate in our multi-D simulations, and follow the subsequent explosion dynamics. By using the turbulent deflagration with DDT model \citep{Seitenzahl2013DDT}, we computed sequences of SN Ia models in the two extremes, where Kelvin-Helmholtz (KH) instabilities are important or not in the ignition phase\footnote{In the multi-dimensional simulation, the KH instabilities down to the resolved scale are consistently modeled.} We show that the models with KH instabilities are mostly near-Chandrasekhar mass WDs. 
These models explode very similarly. Models assuming weak KH instabilities have WDs spanning from subCh-mass to near Ch-mass. Their light curves have the peak $M_{\rm max}$ and with $\Delta M_{15}$ agreeing with the empirical Phillips' relation. The $^{56}$Ni production also spans from $\sim 0.2 - 0.8~M_{\odot}$. These features agree with those in typical SNe Ia surveys, meaning that such a WD could appear like a normal SNe Ia. 

Because the PBH channel does not have a similar mass accretion restriction as the binary channel, a subCh-mass WD can be ignited as long as the PBH mass and WD mass pass the threshold. A direct consequence is that SNe Ia in this channel can occur in the early universe, especially where the WD could be abundantly formed by starburst in galactic haloes. This location also coincides with a high DM density, which further amplifies the occurrence rate. As a result, early SNe Ia could be a result of this channel. The inherent differences between the two subclasses could change the supernova statistics and cosmic metal enrichment process. 

\subsection{Post-Processing Nucleosynthesis}

To capture how the explosion synthesizes chemical elements, in Paper I, we use the post-processing tracers to capture the Lagrangian evolution of the fluid elements in the supernovae \citep{Travaglio2004Tracer}. The tracer scheme places massless points that follow the advection flow of the star passively. They record the thermodynamic history (density, temperature) as a function of time. 

After the hydrodynamics simulations, we pass the thermodynamics history to a general nuclear reaction network \texttt{Torch} \citep{Timmes1999Torch} coupled with the \texttt{Helmholtz} equation of state \citep{Timmes1999Helm}. The network solves the one-zone nucleosynthesis where we choose the 495-isotope network from $^{1}$H to $^{91}$Tc. This network could mostly cover important isotopes that are commonly synthesized in SNe Ia. We use this network to compute the nuclear reactions of the given tracers. 

To explore the metallicity effect, we vary the initial composition by the mass fraction of $^{22}$Ne to represent the metallicity. $^{12}$C and $^{16}$O are chosen to be $1:1$. In \cite{Leung2018Chand} we show that to a very good approximation, the initial metallicity does not play an important role in the explosion dynamics. Thus, the post-process nucleosynthesis is computed based on the hydrodynamics models of fixed metallicity. In some cases, we assume the WD has not experienced the mass accretion phase, which means no simmering phase before the explosion. The initial composition in the metal assumes the same as the solar composition in those models.

\section{Metallicity Dependence of Nucleosynthesis}
\label{sec:results}

\subsection{Models}

\begin{table*}[]
    \label{tab:models}
    \begin{center}
    \caption{The models implemented in this work. $M_{\rm WD}$ is the mass of the PBH, and the radius of the WD is $\rho_{\rm WD, C}$ (in units of gcm$^{-3}$). 
    For each model, the model name is defined as A-B-ZC, where A is the mass coordinate of the first ignition in the progenitor WD in units of $0.1 M_{\odot}$, B stands for the classification in MC19 whether the progenitor assumes KH instabilities, and C stands for the metallicity in solar units.}
    \begin{tabular}{c c c c c c c c c c}
        \hline
         Model & $M_{\rm WD}$ & $\rho_{\rm WD,c}$ & $Z$ & $^{56}$Ni & [Mn/Fe] & [Co/Fe] & [Ni/Fe] & $^{55}$Mn & $^{58}$Ni \\
         \hline
         02B-noKH-Z0 & 0.95 & $2.35 \times 10^7$ & 0 & 0.327 & -1.672 & -1.741 & -0.654 & $7.16 \times 10^{-5}$ & $2.17 \times 10^{-4}$ \\
         02B-noKH-Z0002 & 0.95 & $2.35 \times 10^7$ & 0.002 & 0.322 & -0.817 & -1.731 & -0.617 & $5.08 \times 10^{-4}$ & $2.50 \times 10^{-4}$  \\
         02B-noKH-Z001 & 0.95 & $2.35 \times 10^7$ & 0.01 & 0.308 & -0.411 & -0.685 & -0.317 & $1.26 \times 10^{-3}$ & $4.44 \times 10^{-3}$ \\
         02B-noKH-Z002 & 0.95 & $2.35 \times 10^7$ & 0.02 & 0.295 & -0.250 & -0.522 & -0.084 & $1.80 \times 10^{-3}$ & $1.03 \times 10^{-2}$ \\
         02B-noKH-Z006 & 0.95 & $2.35 \times 10^7$ & 0.06 & 0.257 & 0.037 & -0.380 & 0.348 & $3.41 \times 10^{-3}$ & $3.44 \times 10^{-2}$  \\
         02B-noKH-Z010 & 0.95 & $2.35 \times 10^7$ & 0.10 & 0.223 & 0.219 & -0.429 & 0.553 & $5.09 \times 10^{-3}$ & $5.68 \times 10^{-2}$  \\
         06B-noKH-Z0 & 1.08 & $5.62 \times 10^7$ & 0 & 0.626 & -1.917 & -1.665 & -0.677 & $7.82 \times 10^{-5}$ & $4.14 \times 10^{-4}$  \\
         06B-noKH-Z0002 & 1.08 & $5.62 \times 10^7$ & 0.002 & 0.621 & -1.183 & -1.192 & -0.611 & $4.21 \times 10^{-4}$ & $1.22 \times 10^{-3}$ \\
         06B-noKH-Z001 & 1.08 & $5.62 \times 10^7$ & 0.01 & 0.604 & -0.773 & -0.634 & -0.281 & $1.06 \times 10^{-3}$ & $1.10 \times 10^{-2}$  \\
         06B-noKH-Z002 & 1.08 & $5.62 \times 10^7$ & 0.02 & 0.584 & -0.607 & -0.498 & -0.041 & $1.53 \times 10^{-3}$ & $2.42 \times 10^{-2}$  \\
         06B-noKH-Z006 & 1.08 & $5.62 \times 10^7$ & 0.06 & 0.516 & -0.206 & -0.458 & 0.414 & $3.60 \times 10^{-3}$ & $7.79 \times 10^{-2}$  \\
         06B-noKH-Z010 & 1.08 & $5.62 \times 10^7$ & 0.10 & 0.449 & 0.157 & -0.620 & 0.608 & $7.84 \times 10^{-3}$ & $1.18 \times 10^{-1}$  \\
         10B-noKH-Z0 & 1.26 & $3.11 \times 10^8$ & 0 & 1.090 & -1.043 & -1.052 & -0.455 & $1.02 \times 10^{-3}$ & $1.36 \times 10^{-2}$ \\
         
         10B-noKH-Z0002 & 1.26 & $3.11 \times 10^8$ & 0.002 & 1.084 & -0.948 & -0.969 & -0.394 & $1.26 \times 10^{-3}$ & $1.67 \times 10^{-2}$ \\
         
         10B-noKH-Z001 & 1.26 & $3.11 \times 10^8$ & 0.01 & 1.058 & -0.728 & -0.698 & -0.166 & $2.07 \times 10^{-3}$ & $3.43 \times 10^{-2}$  \\
         
         10B-noKH-Z002 & 1.26 & $3.11 \times 10^8$ & 0.02 & 1.028 & -0.531 & -0.646 & 0.021 & $3.20 \times 10^{-3}$ & $5.65 \times 10^{-2}$  \\
         
         10B-noKH-Z006 & 1.26 & $3.11 \times 10^8$ & 0.06 & 0.908 & 0.038 & -0.799 & 0.377 & $1.10 \times 10^{-2}$ & $1.29 \times 10^{-1}$  \\
         
         10B-noKH-Z010 & 1.26 & $3.11 \times 10^8$ & 0.10 & 0.790 & 0.294 & -0.861 & 0.544 & $1.88 \times 10^{-2}$ & $1.81 \times 10^{-1}$ \\ \hline

         06R-noKH-Z0 & 1.08 & $5.62 \times 10^7$ & 0 & 0.713 & -2.065 & -1.411 & -0.523 & $6.35 \times 10^{-5}$ & $1.23 \times 10^{-3}$  \\
         06R-noKH-Z0002 & 1.08 & $5.62 \times 10^7$ & 0.002 & 0.708 & -1.322 & -1.123 & -0.472 & $3.50 \times 10^{-4}$ & $2.09 \times 10^{-3}$ \\
         06R-noKH-Z001 & 1.08 & $5.62 \times 10^7$ & 0.01 & 0.688 & -0.922 & -0.553 & -0.204 & $8.63 \times 10^{-4}$ & $1.32 \times 10^{-2}$  \\
         06R-noKH-Z002 & 1.08 & $5.62 \times 10^7$ & 0.02 & 0.666 & -0.767 & -0.417 & 0.018 & $1.21 \times 10^{-3}$ & $2.92 \times 10^{-2}$  \\
         06R-noKH-Z006 & 1.08 & $5.62 \times 10^7$ & 0.06 & 0.588 & -0.394 & -0.268 & 0.466 & $2.63 \times 10^{-3}$ & $9.50 \times 10^{-2}$  \\
         06R-noKH-Z010 & 1.08 & $5.62 \times 10^7$ & 0.10 & 0.511 & -0.015 & -0.294 & 0.686 & $5.84 \times 10^{-3}$ & $1.52 \times 10^{-1}$ \\ \hline
         
         06B-noKH-Z0-NS & 1.08 & $5.62 \times 10^7$ & 0 & 0.626 & -1.917 & -1.665 & -0.677 & $7.82 \times 10^{-5}$ & $4.14 \times 10^{-4}$ \\
         06B-noKH-Z0002-NS & 1.08 & $5.62 \times 10^7$ & 0.002 & 0.625 & -1.497 & -1.431 & -0.647 & $2.05 \times 10^{-4}$ & $7.75 \times 10^{-4}$ \\
         06B-noKH-Z001-NS & 1.08 & $5.62 \times 10^7$ & 0.01 & 0.620 & -1.087 & -0.934 & -0.550 & $5.25 \times 10^{-4}$ & $2.24 \times 10^{-3}$  \\
         06B-noKH-Z002-NS & 1.08 & $5.62 \times 10^7$ & 0.02 & 0.615 & -0.912 & -0.663 & -0.429 & $7.82 \times 10^{-4}$ & $5.36 \times 10^{-3}$ \\
         06B-noKH-Z006-NS & 1.08 & $5.62 \times 10^7$ & 0.06 & 0.594 & -0.642 & -0.348 & -0.100 & $1.42 \times 10^{-3}$ & $1.91 \times 10^{-2}$  \\
         06B-noKH-Z010-NS & 1.08 & $5.62 \times 10^7$ & 0.10 & 0.575 & -0.512 & -0.189 & 0.095 & $1.89 \times 10^{-3}$ & $3.33 \times 10^{-2}$ \\ \hline
         
         06B-KH-Z0 & 1.28 & $4.10 \times 10^8$ & 0 & 1.117 & -0.687 & -1.102 & -0.427 & $2.39 \times 10^{-3}$ & $1.66 \times 10^{-2}$ \\
         06B-KH-Z0002 & 1.28 & $4.10 \times 10^8$ & 0.002 & 1.111 & -0.631 & -1.005 & -0.374 & $2.71 \times 10^{-3}$ & $1.95 \times 10^{-2}$ \\
         06B-KH-Z001 & 1.28 & $4.10 \times 10^8$ & 0.01 & 1.085 & -0.473 & -0.761 & -0.168 & $3.84 \times 10^{-3}$ & $3.62 \times 10^{-2}$ \\
         06B-KH-Z002 & 1.28 & $4.10 \times 10^8$ & 0.02 & 1.053 & -0.323 & -0.711 & 0.006 & $5.33 \times 10^{-3}$ & $5.69 \times 10^{-2}$ \\
         06B-KH-Z006 & 1.28 & $4.10 \times 10^8$ & 0.06 & 0.930 & 0.099 & -0.835 & 0.355 & $1.31 \times 10^{-2}$ & $1.27 \times 10^{-1}$ \\
         06B-KH-Z010 & 1.28 & $4.10 \times 10^8$ & 0.10 & 0.809 & 0.322 & -0.922 & 0.524 & $2.08 \times 10^{-2}$ & $1.80 \times 10^{-1}$ \\

          \hline
    \end{tabular}
    \end{center}
\end{table*}

In this work, we have run models that aim at 
extracting the metallicity dependence of the PBH-triggered SNe Ia. 
In Table \ref{tab:models} we list the models computed in this work. The WD mass range- reported in MC19, where the DM can trigger the first nuclear runaway from a subCh-mass 0.95 $M_{\odot}$ up to Ch-mass $\sim 1.3 M_{\odot}$. This corresponds from sub-luminous SNe Ia with a $^{56}$Ni $ = 0.2 M_{\odot}$ to superluminous SNe Ia with $^{56}$Ni $ 1.1 M_{\odot}$. The widespread of $^{56}$Ni can cover the observed diversity of SNe Ia with different peak luminosity.  

The table also reports the final chemical abundances of some key Fe-group elements. Consistent with our previous surveys in SNe Ia \citep{Leung2018Chand, Leung2020SubChand}, a high metallicity will result in a higher $^{55}$Mn and $^{58}$Ni production. It is because the seed $^{22}$Ne can continue its nuclear fusion along the $\alpha$-element chain until it reaches $^{55}$Co, which later decays into $^{55}$Mn by $\beta$-decay.

In this work, we also consider a new series denoted as the ``NS'' series. In the canonical picture of SNe Ia, a WD must experience a certain period of mass accretion from the companion star, especially for the Ch-mass model to reach the critical mass and start the runaway. During the process, the WD experiences the simmering phase where $^{14}$N is slowly converted into $^{22}$Ne. However, for the PBH-triggered SNe Ia, the WD does not necessarily require such accretion. The PBH can trigger the runaway depending on the mass coordinate and the progenitor mass. SubCh-mass WD can be directly formed from a single star. When the PBH ignites the WD, the WD could be young and has not experienced any simmering phase. Its chemical composition besides the C+O rich material remains similar to the spectrum of elements in the solar composition. The NS series reflects this limit for young WDs or WDs in single-star systems. 

\subsection{Metallicity Dependence in Isotopic Production}

\begin{figure*}
    \centering
    \includegraphics[width=0.95\linewidth]{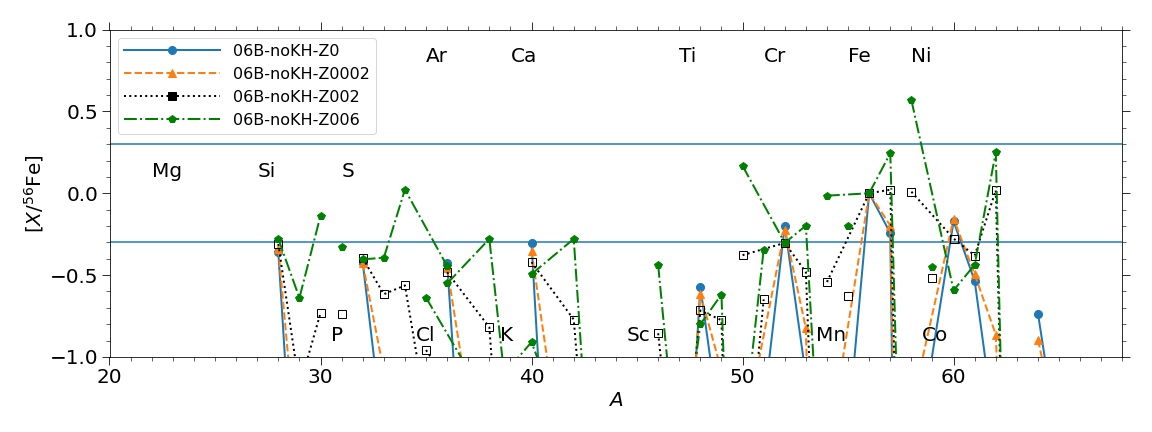}
    \includegraphics[width=0.95\linewidth]{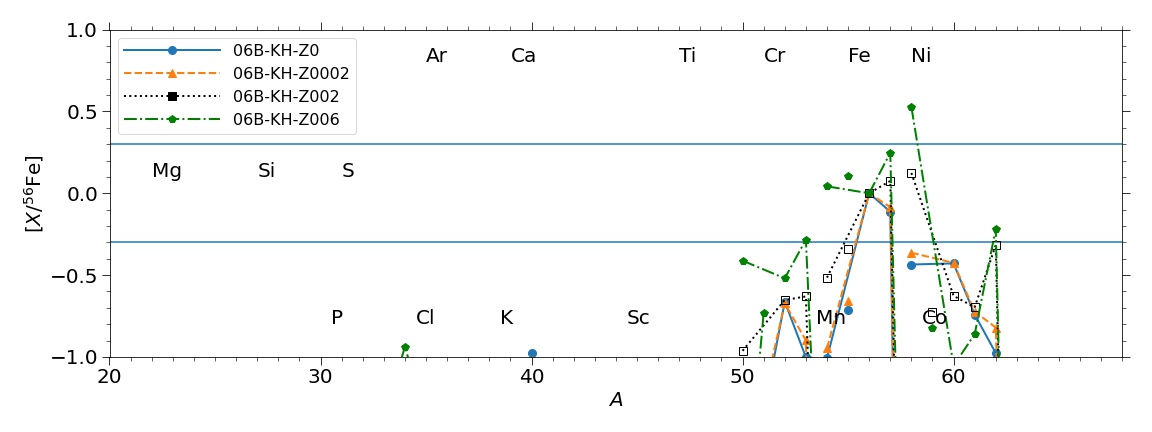}
    
    \caption{(top panel) The isotopic mass fraction ratios [X/$^{56}$Fe] for 06B-noKH-Z0, 06B-noKH-Z0002, 06B-noKH-Z002 and 06B-noKH-Z006. The horizontal lines refer to two times (upper line) and half (lower line) of the solar values. 
    (bottom panel) Same as the top panel but for 06B-KH-Z0, 06B-KH-Z0002, 06B-KH-Z002, 06B-KH-Z006. }
    \label{fig:xiso_Z_plot}
\end{figure*}

In the top panel of Figure \ref{fig:xiso_Z_plot} we plot the isotopic mass fraction ratio [X/$^{56}$Fe]\footnote{[X/$^{56}$Fe] is defined as $\log_{10}$ [(X/$^{56}$Fe)/(X/$^{56}$Fe)$_{\odot}$] such that the zero value corresponds to the same ratio as the solar composition.} for the 06B-noKH series. We remind that this model is chosen as the characteristic model in Paper I because it produces $\sim 0.6~M_{\odot}$ $^{56}$Ni, which is the typical $^{56}$Ni mass observed in actual SNe Ia. We plot the same models but in different metallicity from 0 to 3$Z_{\odot}$. Some isotopes are insensitive to the change of the metallicity, such as most isotopes along the $\alpha$-chain, including $^{28}$Si, $^{32}$S, $^{40}$Ca, $^{48}$Ti, $^{52}$Cr. Neutron-rich isotopes such as $^{30}$Si, $^{34}$S, $^{38}$Ar, $^{50}$Cr, $^{58}$Ni, sharply increase with the metallicity.

In the bottom panel, we plot similar to the top panel but for the 06B-KH series. The model explodes close to a pure detonation, where most material is incinerated into Fe-group elements. Most of the Si-group elements cannot be found in the noKH counterpart. Despite that, the metallicity dependence in this series is still visible for isotopes including $^{55}$Mn, $^{54, 57}$Fe, $^{58, 62}$Ni. No observable amount of isotope beyond $^{62}$Ni is found. 

\subsection{Metallicity Dependence in Elemental Production}

\begin{figure*}
    \centering
    \includegraphics[width=0.48\linewidth]{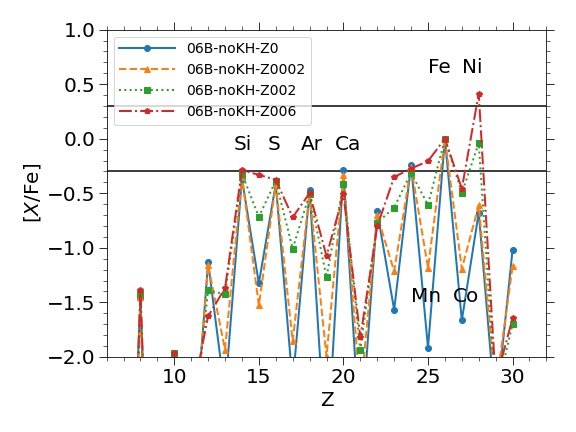}
    \includegraphics[width=0.48\linewidth]{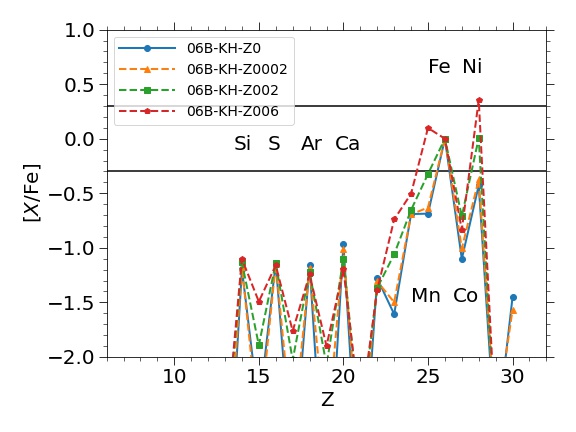}
    
    \caption{(left panel) The elemental mass fraction ratios [X/Fe] for 06B-noKH-Z0, 06B-noKH-Z0002, 06B-noKH-Z002 and 06B-noKH-Z006. The horizontal lines refer to two times (upper line) and half (lower line) of the solar values. 
    (right panel) Same as the top panel but for 06B-KH-Z0, 06B-KH-Z0002, 06B-KH-Z002, 06B-KH-Z006. }
    \label{fig:xele_Z_plot}
\end{figure*}

To further characterize the yield of the PBH-triggered SNe Ia, we plot in Figure \ref{fig:xele_Z_plot} the elemental abundance of the same set of models as in the previous plot.

In the noKH series, the $\alpha$-chain elements are less sensitive to the metallicity except for Ni. It is because they are synthesized by the detonation, where the weak interaction is less effective. On the other hand, Ni is found in both deflagration and detonation ash. The dependence is more clearly seen for the odd-number elements in both the Si-group (P, Cl, K) and the Fe-group elements (V, Mn, Co). A higher metallicity favours the production of these elements. In all models, only Ni can be matched with the solar composition. Other elements show underproduction regardless of the initial metallicity. 

In the KH series, the metallicity dependence is more subtle because of the suppressed production of Si-group elements. The dependence is reflected by the trend for Fe-group elements from Sc to Ni, and a higher metallicity favours their production. Ni is the only element that can be compatible with solar abundance. 

\subsection{Applications to SN Ia Light Curves and Remnants}

\begin{figure*}
    \centering
    \includegraphics[width=0.48\linewidth]{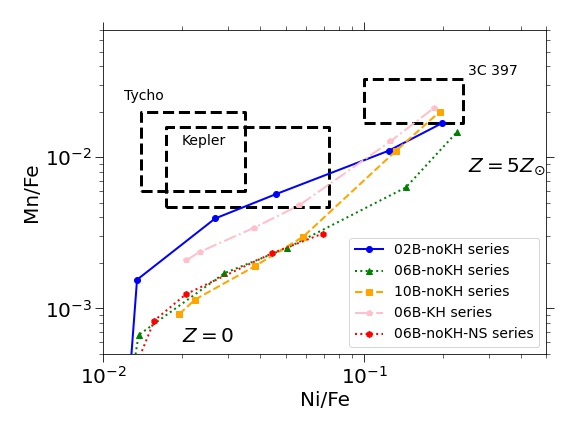}
    \includegraphics[width=0.48\linewidth]{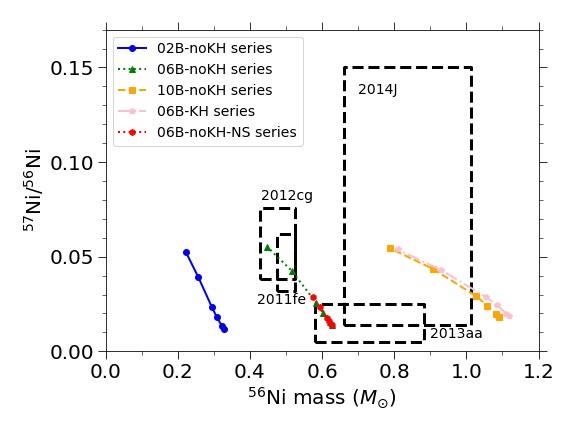}
    
    \caption{(top panel) The mass fraction ratios Mn/Fe against Ni/Fe for models from Table \ref{tab:models}. Each line corresponds to the same models but with $Z = 0, 2\times 10^{-3}, 0.01, 0.02, 0.06, 0.10$ respectively. We also include the series for 06B-noKH-NS to illustrate the WDs without simmering. The observational data from supernovae remnants Kepler \citep{Yamaguchi2014Kepler}, Tycho \citep{Park2013Tycho}, and 3C 397 \citep{Yamaguchi20153C397} are included as boxes corresponding to the 1 $\sigma$ level. 
    (bottom panel) The mass fraction $^{57}$Ni/$^{56}$Ni against $^{56}$Ni mass for models from Table \ref{tab:models} with the same metallicity points as the top panel. The rectangular boxes correspond to the observed range from SN 2011fe \citep{Tucker2022SN2011fe}, SN 2012cg \citep{Graur2016SN2012cgLateLC}, SN 2013aa \citep{JacobsonGalan2018SN2013aa}, and SN 2014J \citep{Graur2018SN2015F}. For supernovae with multiple constraints, the boxes center at the average of these measurements, and with boundaries defined as the minimum and maximum of these measurements.}
    \label{fig:snr_compare}
\end{figure*}

To test the chemical yields from the PBH-triggered SN Ia models, we compare their elemental production with some well-observed SNe Ia in the literature. That includes the chemical composition derived from the X-ray spectra, the supernova remnants, and the late-time light curve of some nearby SNe Ia. For the numerical reference, in Appendix \ref{sec:appA} we show the yield table for the characteristic SN Ia model 06B-noKH for various metallicity, and in Appendix \ref{sec:appB} the comparison of the same model with the canonical subCh-mass model SK18.  

In the left panel of Figure \ref{fig:snr_compare}, we compare the Mn/Fe against Ni/Fe for our models compared to other SN Ia remnants, including Tycho \citep{Park2013Tycho}, Kepler \citep{Yamaguchi2014Kepler}, and 3C 397 \citep{Yamaguchi20153C397}. The 3C 397, in particular, is a strong candidate that supports a near Ch-mass and high metallicity progenitor due to its exceptionally high Mn and Ni mass ratios. All model series show homogeneous variation against metallicity: A high metallicity enhances the production of Mn and Ni relative to Fe. The NS series mimics the original series except that its effective metallicity corresponds to a much lower metallicity model in the original 06B-noKH series. Note that the high metallicity might be in contrast with the typical GCE model, where a maximum of [Fe/H]$\sim0.3-0.5$ is observed. A more recent analysis of the elemental abundance suggests a lower value of these Fe-group elements \citep{Ohshiro20213C397}, suggesting a lower metallicity model. The recently launched XRISM (X-ray Imagine and Spectroscopy Mission) Telescope will provide tighter constraints for revisiting these SNRs.

The Kepler SNR can be explained by near-solar metallicity models of the 02B-noKH-series or 06B-KH-series. On the other hand, the observed limits of SNR 3C 397 can be approached by the extremely high metallicity models in all series except 06B-noKH-NS. This gives a dual interpretation that these two SNRs have either a Ch-mass or subCh-mass WD progenitor. No model approaches the Tycho SNR in the parameter space, suggesting that it could be more consistently explained by standard SN Ia models. 

In the right panel, we compare the isotopic ratios for $^{57}$Ni and $^{56}$Ni. The $^{56}$Ni can be well calibrated from the light curve, especially the peak luminosity, as noted in Arnett's law \citep{Arnett1982LC}. The $^{57}$Ni can be derived from the late-time decay rate of the light curves. A high metallicity model has a lower $^{56}$Ni but higher $^{57}$Ni. The parameter space, especially $^{56}$Ni, can uniquely distinguish the progenitor sequence.

Most observed SN Ia LCs can be mapped to one or multiple PBH-triggered SN Ia models. The mass ratio of the isotope pair for SN 2011fe \citep{Tucker2022SN2011fe}, 2012cg \citep{Graur2016SN2012cgLateLC}, and 2013aa \citep{JacobsonGalan2018SN2013aa} can be explained by the 06B-noKH(-NS) series. This suggests their subCh-mass WD as a common interpretation of these SNe Ia with a mass $\sim 1.0-1.1~M_{\odot}$. On the other hand, the SN 2014J \citep{Graur2018SN2015F} can be correlated with the explosion of a high metallicity model of either 10B-no or 06B-KH series. Notice that these two series correspond to the near Ch-mass WDs. The low-mass 02B-noKH series does not match any SNe Ia observed.

\section{Influence on Galactic Chemical Evolution}
\label{sec:GCE}

\subsection{Governing Equations}

We use the GCE code reported in \cite{Timmes1995GCE}. The code sums the contribution of massive stars, low-mass stars, Type Ia supernovae, and novae to study how generations of stars and supernovae cumulatively change the metal content on the galactic scale, by the governing equation: 
\begin{equation}
\begin{aligned}
\frac{d\sigma_i}{dt} &= 
\int_{0.8}^{3} B(t - \tau(m)) \, \Psi(m) \times X_i(t - \tau(m)) \, dm  \\
&\quad + C \int_{3}^{16} \int_{\mu_m}^{0.5} B(t - \tau(m)) \, \Psi(m) \, f(\mu) \\
&\quad \times X_i(t - \tau(m)) \, d\mu \, dm \\
&\quad + (1 - C) \int_{3}^{16} B(t - \tau(m)) \, \Psi(m)  \\
&\quad \times X_i(t - \tau(m)) \, dm \\
&\quad + \int_{11}^{40} B(t - \tau(m)) \, \Psi(m) \times X_i(t - \tau(m)) \, dm \\
&\quad + (1 - C) \, D_{\rm PBH} \int_{3}^{8} B(t - \tau(m)) \, \Psi(m) \\
&\quad \times X_i(t - \tau(m)) \, dm \\
&\quad - B(t) \frac{\sigma_i}{\sigma_{\rm gas}} + \dot{\sigma}_{i,\rm gas} + \frac{\sigma_i}{\tau_{1/2}} M_\odot \, \text{pc}^{-2} \, \text{Gyr}^{-1}
\end{aligned}
\label{eq:GCE}
\end{equation}
We remind that $d\sigma_i/dt$ is the time rate of change of surface mass density of the $i$-th isotope, such that the important galactic processes include the stellar death, stellar birth, gas inflow and outflow, and radioactive decay. $B(t)$ represents the birthrate function at time $t$. $\tau(m)$ stands for different lifetimes of main-sequence stars at mass $m$ and $\Psi(m)$ is the initial mass function. The term $X_i(t - \tau(m))$ represents the abundance of an isotope in stars formed at time $t - \tau(m)$. The free parameter $C$ is the fraction of intermediate-mass stars in binary systems that produce Type Ia supernovae, with $f(\mu)$ describing the distribution of binary mass ratios $\mu$. 
The last line contains all the source terms for the stellar birth, galactic in- and outflow, and the changes due to radioactive decay. 
The newest addition is our PBH-triggered SN Ia channel (second-last line). The new parameter $D_{\rm PBH}$ is the fraction of WDs that undergo PBH-induced SNe Ia. We integrate stars in the $3-8\,M_\odot$ mass range that are not part of Type Ia binaries but may instead explode due to interactions with PBH. 

For the stellar input, the code assumes the Sa;peter relation for the Initial Mass Function (IMF) and the Schmidt's Star formation function with a dependence on the total mass was adopted \citep{Schmidt1959SF1, Schmidt1963SF, Lacey1985SF}. The gas surface density depends on the location, which includes the disk with a exponentially decreasing profile, and the bulge with an inverse square profile \citep{Binney1987GacDyn}. In Table \ref{tab:parameters} we tabulate the parameters, their physical meaning and the ranges used in this work. We refer interested readers to the documentation paper \citep{Timmes1995GCE} for the detailed implementation and testing.

\begin{table*}
    \centering
    \caption{The physical parameters used in the GCE simulations with their corresponding equations and values or ranges used in this work.}
    \begin{tabular}{c c c c}
    \hline
     Parameter  & Physical Meaning & Equation & Value \\ \hline
    $\nu$ & Schmidt star formation efficiency factor & $B = \nu \sigma_{\rm tot} (\sigma_{\rm gas}/\sigma_{\rm tot})^n$ & 2.2 \\ 
    $n$ & Schmidt star formation exponent & same as above & 2.0 \\
    $s$ & Salpeter exponent in IMF & $\Psi(m) \sim m^s$ & -1.35 \\
    $m_{\rm IMF,lo}$ & Lower limit of IMF ($M_{\odot}$) & Eq. (1) & 0.08 \\
    $m_{\rm IMF,hi}$ & Lower limit of IMF ($M_{\odot}$) & Eq. (1) & 40.0 \\
    $r_{\rm zone}$ & Distance of zone (solar neighbourhood) (kpc) & $\sigma_{\rm tot,disk} = K_{\rm disk} e^{-r/r_{\rm disk}}$ & 8.5 \\
    $K_{\rm disk}$ & Solar vicinity surface density ($M_{\odot}$ pc$^{-2}$) & & 75.0 \\
    $K_{\rm bulge}$ & Galactic center surface density ($M_{\odot}$ pc$^{-2}$) & $\sigma_{\rm tot,bulge} = K_{\rm bulge} / (r + r_{\rm bulge})^2$ & $1.0 \times 10^4$ \\ 
    $r_{\rm disk}$ & Extent of inverse square profile (kpc) & $r_{\rm disk} = (2 + r_{\rm bulge}) / 2$ & 2.0 \\
    $\tau_{\rm disk}$ & Timescale for disk formation (Gyr) & $\dot{\sigma}_{\rm tot} = A(r) e^{-t/\tau_{\rm disk}}$ & 4.0 \\ 
    $m_{\rm II,lo}$ & Lower limit for Type II rates & Eq. (1) & 11.0 \\
    $m_{\rm II,lo}$ & Upper limit for Type II rates & Eq. (1) & 40.0 \\
    $m_{\rm I,lo}$ & Lower limit of binary systems & Eq. (1) & 3.0 \\
    $m_{\rm I,hi}$ & Lower limit of binary systems & Eq. (1) & 16.0 \\
    $C$ & Type Ia amplitude factor & Eq. (1) & (0.0 - 0.030) \\
    $D_{\rm PBH}$ & PBH-channel amplitude factor & Eq. (1) & (0.0 - 0.20) \\
    $f_{\rm single}$ & fraction of single star in PBH & Eq. (1) & (0.0 - 1.00)\\
        \hline
    \end{tabular}
    \label{tab:parameters}
\end{table*}

\subsection{Models}

\begin{table*}[]
    \centering
    \caption{The models used for Galactic Chemical Evolution and the best-fit parameters from the parameter surveys presented in this artice. Columns ``Massive Star'' and ``SN Ia'' stand for the massive star models and SNe Ia in the binary channel models. The PBH stands for the explosion channel presented in Table \ref{tab:models}. Parameters $f{_PBH}$, $f_{\rm single}$, and $t_{\rm off}$ are the best-fit parameters in the parameter survey for each set of supernova model. $\chi^2$ is the corresponding minimum $\chi^2$. }
    \begin{tabular}{c c c c c c c c c c c c c}
        \hline
         Model & Massive Star & SN Ia & PBH & $D_{\rm PBH}$ & $f_{\rm single}$ & $t_{\rm off}$ & $\chi^2$ & [Mn/Fe] & [Ni/Fe] & Others \\ \hline
         L25-LN18(Ka4)-BKH & L25 & LN18 & B-KH & 0.0130 & 0.87 & 9.0 & 45.64 & -0.196 & 0.114 \\ 
         L25-LN18(Ka4)-BnoKH & L25 & LN18 & B-noKH & 0.0198 & 0.56 & 9.0 & 45.10 & -0.196 & 0.112 \\
         L25-LN18(Ka4)-RnoKH & L25 & LN18 & R-noKH & 0.0198 & 0.67 & 9.0 & 46.82 & -0.196 & 0.112 \\
         L25-SK18-BKH & L25 & SK18 & B-KH & 0.0102 & 0.83 & 9.0 & 43.72 & -0.140 & -0.023 \\
         L25-SK18-BnoKH & L25 & SK18 & B-noKH & 0.0198 & 0.74 & 9.0 & 42.83 & -0.139 & -0.023 \\
         L25-SK18-RnoKH & L25 & SK18 & R-noKH & 0.0198 & 0.99 & 9.0 & 42.39 & -0.140 & -0.023 \\ \hline
         L25-LN18(Ka4)-BnoKH-NS & L25 & LN18 & B-noKH & 0.0198 & 0.59 & 9.0 & 40.21 & -0.197 & 0.110 &  No simmering  \\
         L25-SK18-BnoKH-NS & L25 & LN18 & B-noKH & 0.0198 & 0.76 & 9.0 & 38.66 & -0.140 & -0.026 & No simmering \\
        \hline         
    \end{tabular}

    \label{tab:GCE_model}
\end{table*}

In Table \ref{tab:GCE_model} we list the GCE parameter survey performed in this article. Each parameter survey takes the input of massive star models (taken from \cite{Leung2025Perseus1} for stars from 15 -- 40 $M_{\odot}$, SNe Ia \citep{Leung2018Chand,Shen2018DDDDDD}, and the PBH-triggered SNe Ia with different ignition kernels as stellar inputs.  

We perform the parameter survey for each combination of massive star models, SNe Ia models for the binary channel, and the PBH channel. In each model, we obtain the time evolution of the chemical elements. Then we compare the chemical abundances from 6 elements (Si, S, Ar, Ca, Mn, and Ni) with the stellar abundance data from the SAGA database \citep{SAGA2008}. Given the vast amount of stellar abundance data, we group the data in six bins of metallicity from [Fe/H] $\in (-3.0, 0.0)$ and compute their average and standard deviation $\sigma_{\rm SAGA,i}(Z_n)$. We compare the $\chi^2$-square for the $i$-th element by 
\begin{equation}
    \chi^2_i = \sum_n \frac{(X_i(Z_n) - X_{\rm SAGA,i}(Z_n))^2}{\sigma_{\rm SAGA,i}(Z_n)}.
\end{equation}
Then we collect these $\chi^2$ for the total $chi^2$ as an overall measurement of the ``fit'' of the model with the stellar abundance. 

In this article, we focus on four different types of PBH models: (1) BnoKH series, (2) RnoKH series, (3) BKH series, and (4) BnoKH-NS series. They span the possibilities of the PBH channel with contrasting initial flame kernels, or contrasting initial progenitor WD (the noKH series contains primarily subCh-mass WDs, while those in the KH series are primarily Ch-mass WDs), and contrasting scenarios (the NS series corresponds to WDs in single star systems where the WDs have not experienced the simmering phase).

Each parameter survey consists of sampling different $D_{\rm PBH}$, the fraction of single WD ($f_{\rm single}$), and the switch-off time. $D_{\rm PBH} \in (0, 0.02)$ refers to the fraction of WD that can undergo as PBH-triggered SNe Ia. The variable $f_{\rm single} \in (0, 1)$ corresponds to the fraction of subCh-mass WD progenitor within the PBH population. Note that for WD with an initial mass above 1.1 $M_{\odot}$, they are unlikely to be formed by a single star, and hence they are produced in a binary system. This parameter represents the fraction of PBH-triggered SNe Ia evolved from single-star vs. binary-star systems. 

We also include the parameter switch-off time to control the relative importance of the PBH channel in the mid-age universe (redshift $\sim 2-4$). The switch-off time is the moment when the PBH channel is zero, leaving only SNe Ia from the binary channel. We use this in a later section to highlight how the PBH affects the abundance pattern of the recent universe. 
In general the event rate triggered by this channel is proportional to local PBH and the WD density. This mimics the effects when DM-induced channel becomes less important locally.

\subsection{Parameter Survey}

\begin{figure*}
    \centering
    \includegraphics[width=0.48 \textwidth]{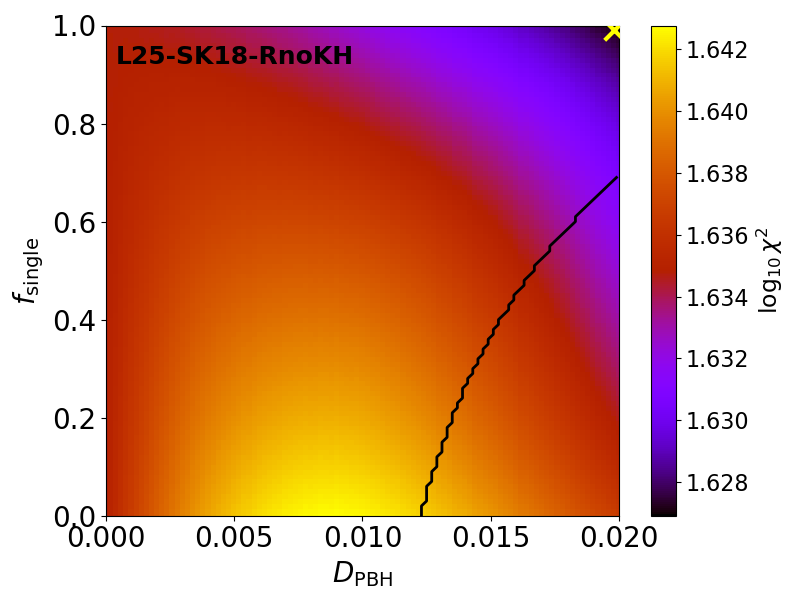}
    \includegraphics[width=0.48 \textwidth]
    {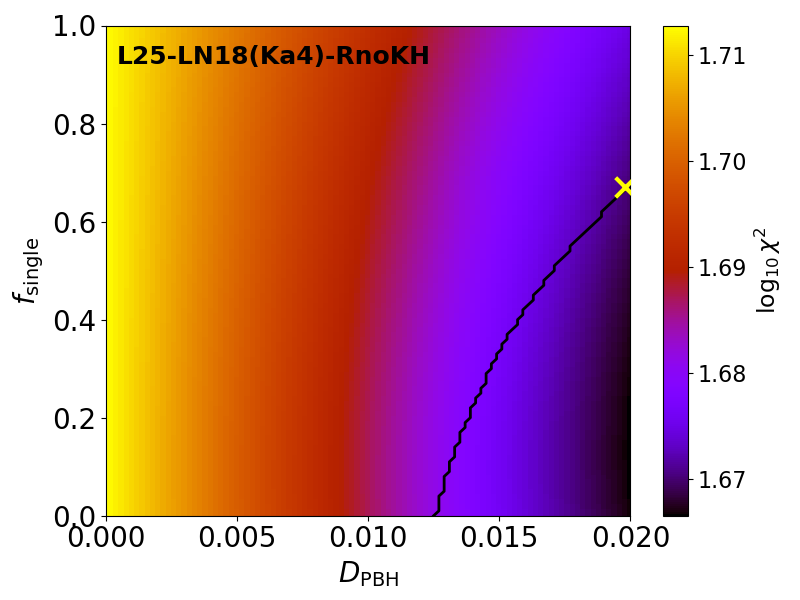}

    \includegraphics[width=0.48 \textwidth]
    {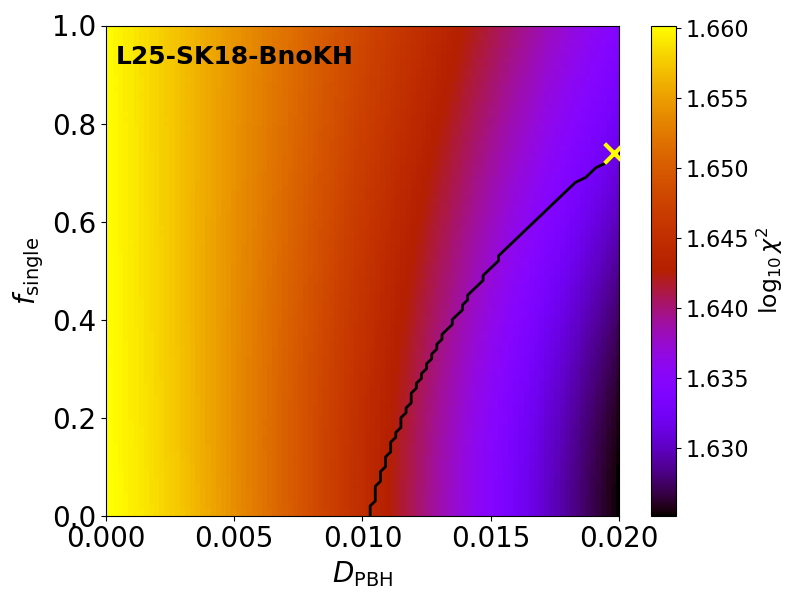}
    \includegraphics[width=0.48 \textwidth]{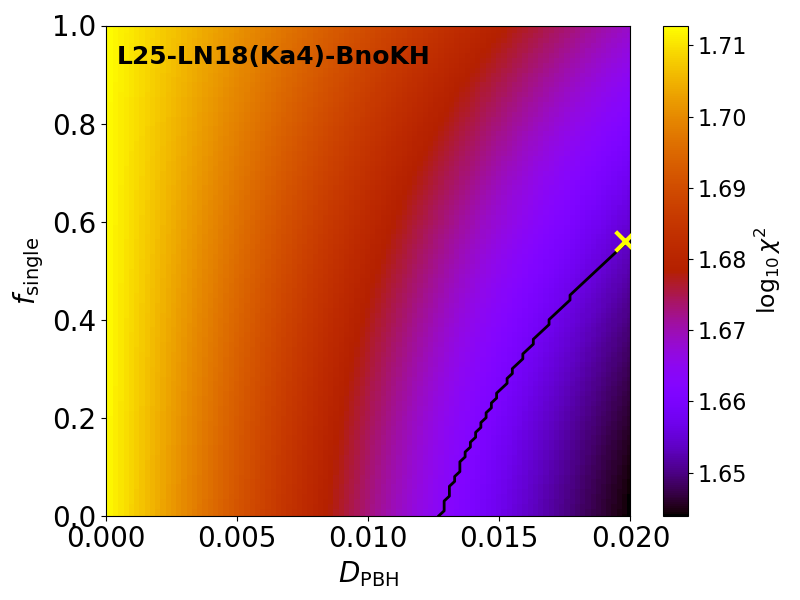}
    
    \includegraphics[width=0.48 \textwidth]
    {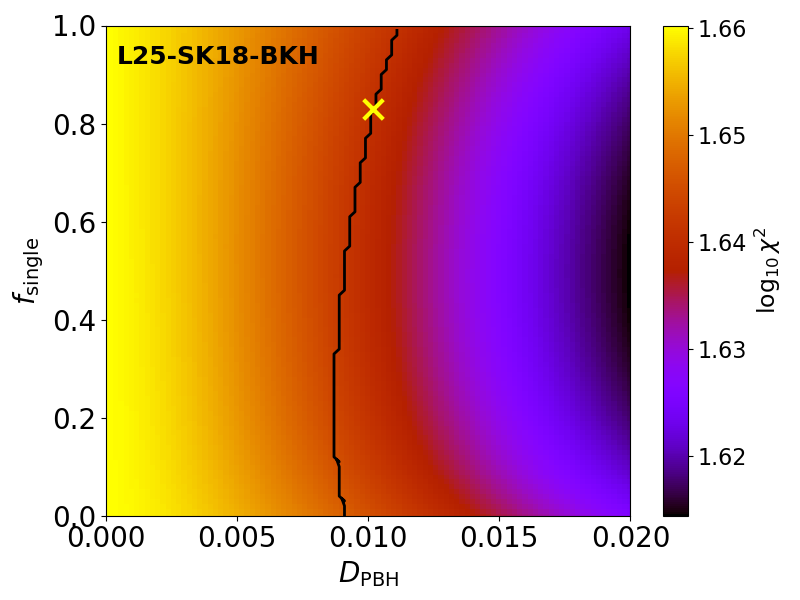}
    \includegraphics[width=0.48 \textwidth]
    {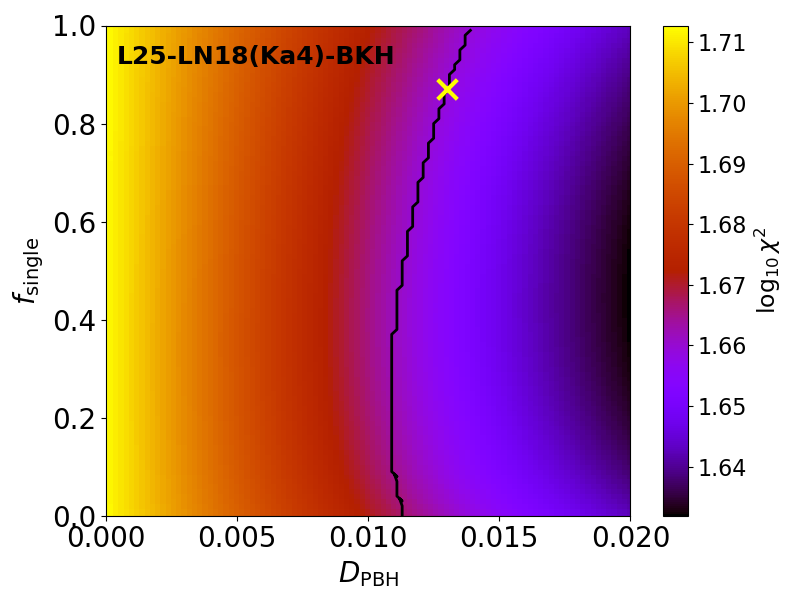}
    \caption{The left column shows the GCE model using the SK18 SN Ia model including L25-SK18-RnoKH (top left panel), L25-SK18-BnoKH (middle left panel), and L25-SK18-BKH (bottom left panel). The right column shows the LN18(Ka4) Type Ia model for L25-LN18(Ka4)-RnoKH (top right panel), L25-LN18(Ka4)-BnoKH (middle right panel), and L25-LN18(Ka4)-BKH (bottom right panel). The black line corresponds to the boundary where, on the right, the models do not exhibit a one-to-one mapping between metallicity and age, which we exclude from this parameter space. The yellow cross marks the parameters, \( D_{\mathrm{PBH}} \) and \( f_{\mathrm{single}} \), for the overall best-fit model.}
    \label{fig:GCE_chisq}
\end{figure*}

In Figure \ref{fig:GCE_chisq} we plot the colour-plot for the $\chi^2$ results for different $D_{\rm PBH}$ and $f_{\rm single}$ used in the parameter survey. The black line in each plot corresponds to the boundary where, on the right-hand side of the boundary, the PBH channel creates too many SNe Ia in the early universe, where the [Fe/H] does not monotonically increase with time. We exclude this portion of parameter space when we search for the minimum configuration. 

In all models, the $D_{\rm PBH} = 0$ is not the configuration with the minimum $\chi^2$. This suggests that the PBH channel cannot be excluded, as it makes a better fit of the chemical elements trends. However, the best value depends strongly on the choice of the PBH-triggered SN Ia models, which could range between $0.01-0.02$. The $f_{\rm single}$ is relatively less important in most cases. 

\subsection{Best-fit Models}


\begin{figure*}
\centering
\includegraphics[width=0.48 \textwidth]{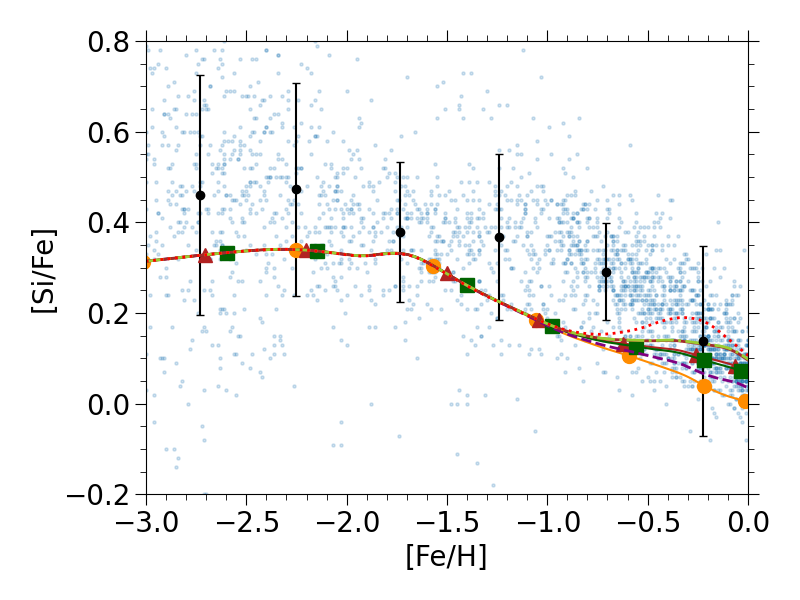}
    \includegraphics[width=0.48 \textwidth]{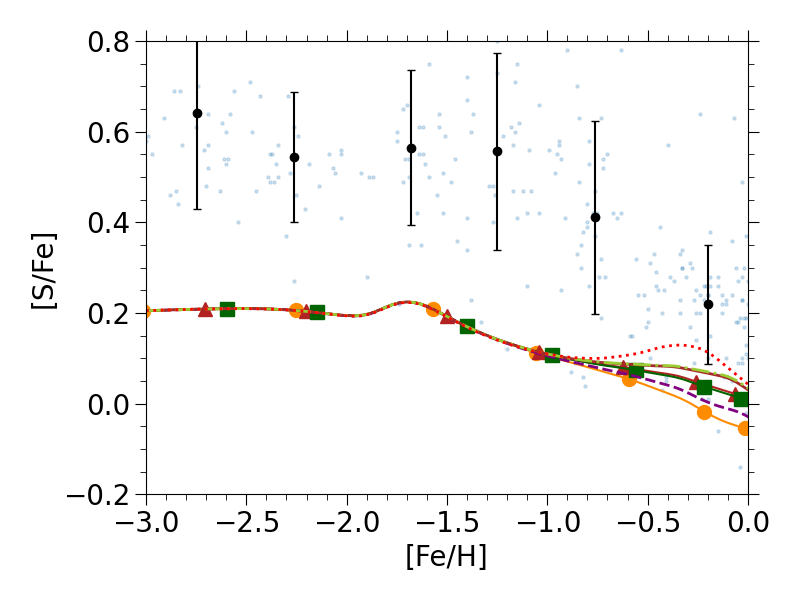}
    \includegraphics[width=0.48 \textwidth]{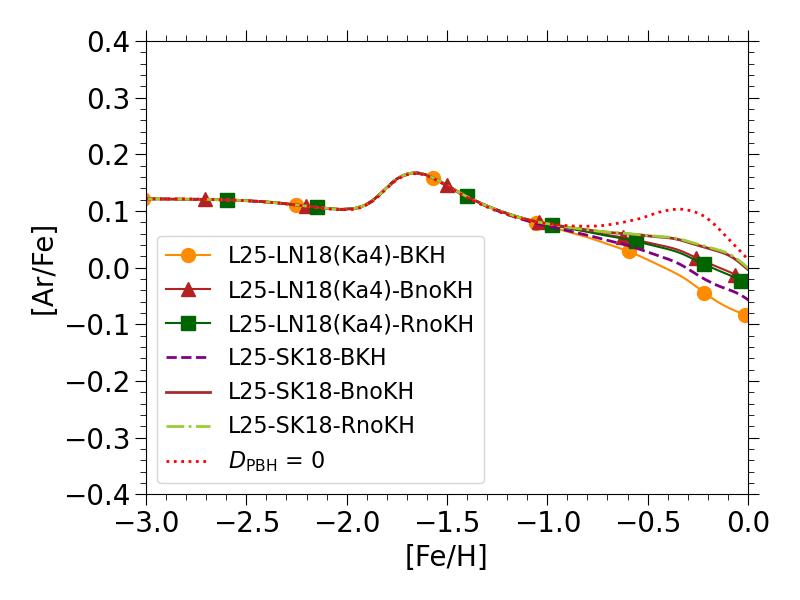}
    \includegraphics[width=0.48 \textwidth]{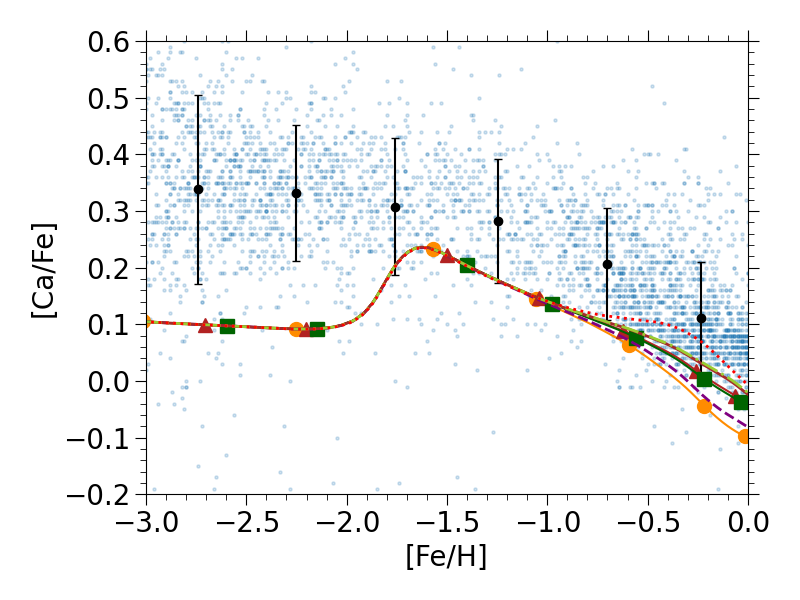}
    \includegraphics[width=0.48 \textwidth]{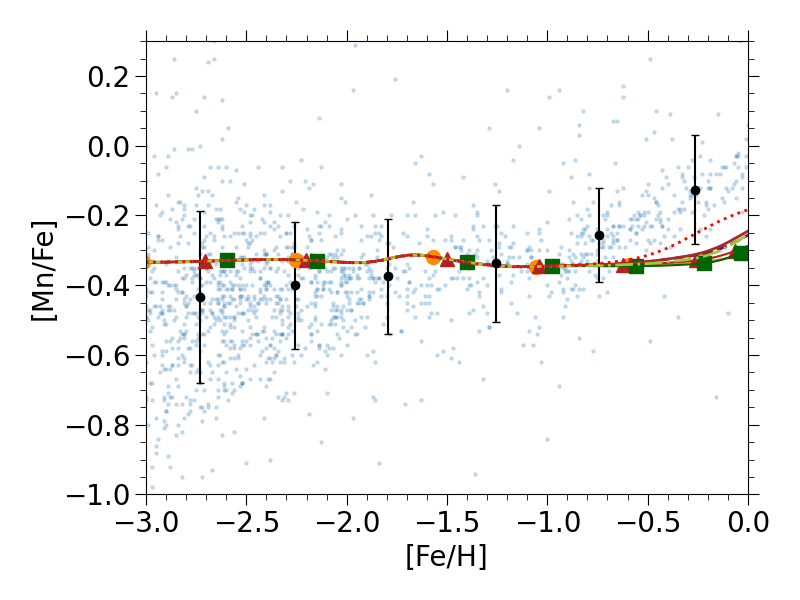}
    \includegraphics[width=0.48 \textwidth]{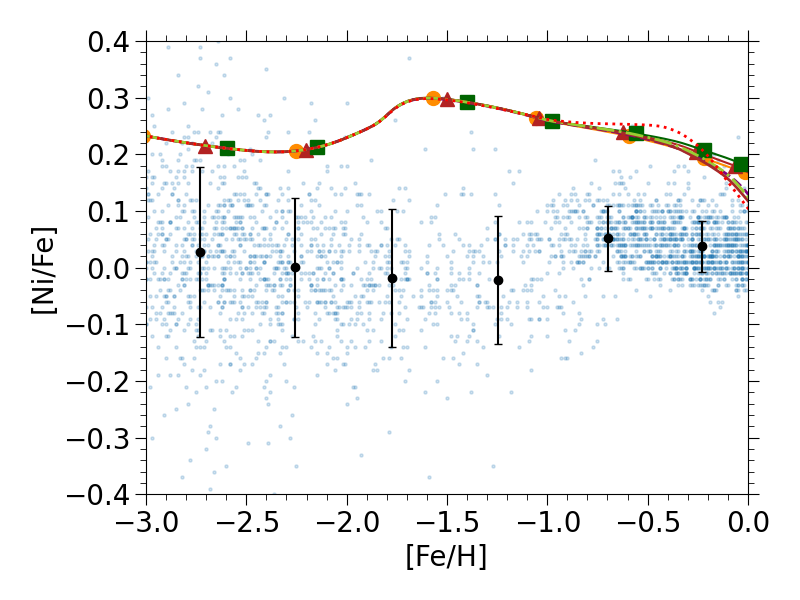}

\caption{(top left panel) The trend of [Si/Fe] against [Fe/H] from Galactic chemical evolution models for the best-fit version of each model, with optimal values of \( D_{\mathrm{PBH}} \) and \( f_{\mathrm{single}} \) applied (See Table \ref{tab:GCE_model}). 
The models are: L25-LN18(Ka4)-BKH (dark orange line with circles), L25-LN18(Ka4)-BnoKH (firebrick line with triangles), L25-LN18(Ka4)-RnoKH (dark green line with squares), L25-SK18-BKH (purple dashed line), L25-SK18-BnoKH (brown line), and L25-SK18-RnoKH (yellow-green dash–dot line). 
Black points with error bars represent binned observational data from the SAGA database, and small blue points represent star measurements. The dotted line refers to the case $D_{\rm PBH} = 0$ with L25-SK18. Other ratios displayed are [S/Fe] (top right panel), [Ar/Fe] (middle left panel), [Ca/Fe] (middle right panel), [Mn/Fe] (bottom left panel), and [Ni/Fe] (bottom right panel).}
\label{fig:GCE_elements}
\end{figure*}

In Figure \ref{fig:GCE_elements} we plot the elemental trends for the best models of the 6-parameter surveys of this article. We also include the stellar data, their metallicity-binned average, and standard deviation to guide the vision. We also include the case $D_{\rm PBH}=0$ for the model L25-SK18-RnoKH to highlight the impact of the PBH models.

Si has a fairly constant trend in the data up to [Fe/H] $\sim -1$ from a super-solar value steadily to the solar value at solar metallicity. The best-fit models have similar trends, but the transition is earlier. Different PBH models only affect the dispersion close to [Fe/H] $\sim 0$. However, we note that the ``knee'' of our model occurs at [Fe/H]$\sim-1.5$ while the stellar data shows a later bend. This is related to the delayed time distribution (DTD) where binary systems have a systematic delay time for the explosion to take place \citep{Greggio2005DTD} We further discuss the role of the delayed time distribution in Section \ref{sec:caveats}.

For S, all models are distinct from the stellar data. Early stars exhibit a high S/Fe ratio, which is not seen in any SN Ia models in both the binary and PBH channels. It suggests that such an early trend can be a result of massive star models or other stellar sources that are not included in the current GCE code. Despite that, the shape of the curve can mimic both the plateau phase and the falling phase of this element. The PBH models also show their effect in high metallicity environments only. 

For Ar, there is no chemical abundance data from stars in the SAGA database to compare with. The models only show differences in high metallicity environments similar to Si and S. Given the similarity of all best-fit models, it will require very precise measurement of this element, such that additional constraints can be cast on the PBH models. 

For Ca, the low metallicity stars have, in general, higher abundance than the current model, suggesting a higher input from the massive stars is necessary. This could also be attributed to more exotic PBH models where a thick He envelope is involved. The He-detonation at low metallicity can produce super-solar Ca/Fe in the ejecta. Beyond [Fe/H] $> -1.7$, the falling trend of the model agrees with the stellar data. The PBH models increase the dispersion but at a slope consistent with the stellar data. 

For Mn, the stellar data is flat up to [Fe/H] $\sim -1$ and then gradually increases. The best-fit models also show a similar plateau phase, but the rising trend is suppressed. This is because in the parameter survey, in order to minimize the $\chi^2$, the code favours the models by lowering Ni. Since SNe Ia produce supersolar Ni/Fe, the code tends to reduce the importance of SNe Ia. This results in a mild increase in the [Mn/Fe]. The PBH models bring almost no change to the evolution, unlike other elements. 

Ni is one of the tricky elements. The abundance pattern from the SAGA database shows an almost uniform value across cosmic history. This means that the production by massive stars and the two SN Ia channels have to be very similar in the first place. The high value in all our best-fit models shows that the massive stars overproduce this element by 0.2 dex. There is a small rising trend in our model at [Fe/H] = -1.5, which also appears in the observational data, but at a different metallicity [Fe/H] = -1.0. To further fit this parameter, it requires fine-tuning of the switch-off time, such that PBH will not over-accelerate the Fe synthesis in the early universe. The PBH model cannot account for the diversity of Ni/Fe measurements in low metallicity stars, but so for near $Z_{\odot}$ stars. 

By comparing the models with the case of $D_{\rm PBH}=0$, the PBH model tends to change the evolution near [Fe/H]$\sim-0.5$. The additional PBh channel suppresses the bumps in the creation of Si-group elements by about 0.1--0.2 dex, and the average stellar trends do not exhibit such bumps. On the other hand, it leads to an earlier rise (due to Ch-mass model) in [Mn/Fe] at the same metallicity. Meanwhile, the bump in [Ni/Fe] leads to a greater offset.

We remind that these models assume $t_{\rm off}=9$ Gyr. This means that the PBH channel tends to affect only the older population stars and the newer population is only shaped by recent SNe Ia by the binary channel. While it could be counterintuitive that the PBH affects the elemental evolution starting from [Fe/H]$>-1$, notice that WDs are evolved from lower-mass stars ($3-8~M_{\odot})$. By the time these WDs explode, a number of CCSNe have already exploded, where the local metallicity has raised to $\sim-1$. Therefore, the stars with [Fe/H]$>-1$ also correspond to the stars within the first 9 Gyr.

\subsection{Effects of PBH-induced SN Ia on Mn production}

\begin{figure}
    \centering
    \includegraphics[width=0.95\linewidth]{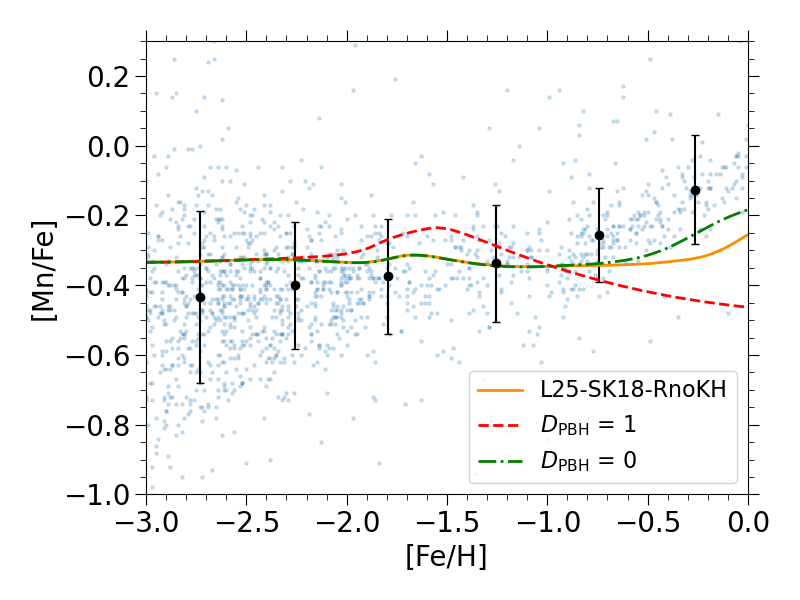}
    \includegraphics[width=0.95\linewidth]{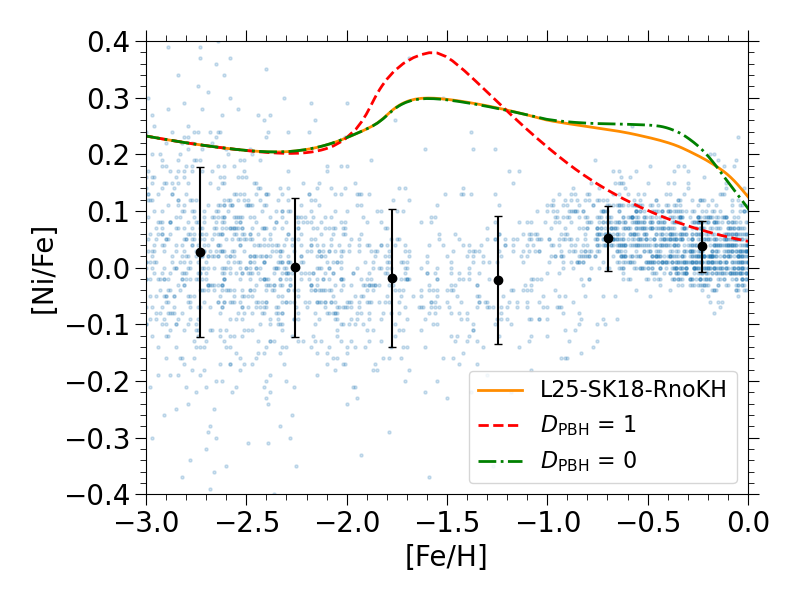}
    \caption{(top panel) [Mn/Fe] versus [Fe/H] for the best model, L25-SK18-RnoKH (orange line), against a similar model in which the universe contains no PBH-induced SNe~Ia (\( D_{\mathrm{PBH}} = 0 \)) (green dash-dotted line), and a similar model in which the universe contains only PBH-induced SNe~Ia (\( D_{\mathrm{PBH}} = 1 \)) (red dashed line). Black points with error bars represent binned observational data from the SAGA database, and small blue points represent star measurements.
    (bottom panel) Same as the top panel but for [Ni/Fe].}
    \label{fig:GCE_dpbh}
\end{figure}

\begin{figure}
    \centering
    \includegraphics[width=0.95\linewidth]{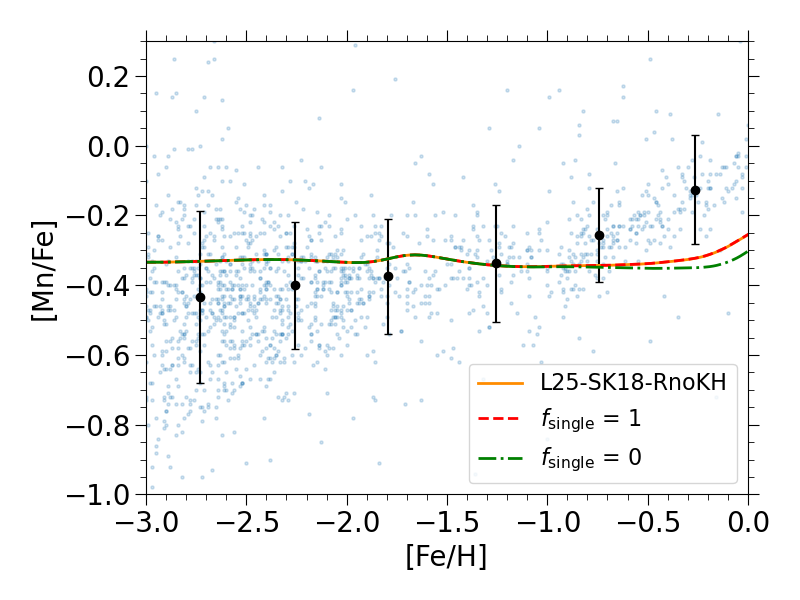}
    \caption{[Mn/Fe] versus [Fe/H] for the best model, L25-SK18-RnoKH (orange line), is compared with a similar model in which the fraction of single-WD star systems (\( f_{\mathrm{single}} \)) is 0 (green dash-dotted line) and a similar model in which \( f_{\mathrm{single}} = 1 \) (red dashed line). Black points with error bars represent binned observational data from the SAGA database, and small blue points represent star measurements.}
    \label{fig:GCE_fsing}
\end{figure}

\begin{figure}
    \centering
    \includegraphics[width=0.95\linewidth]{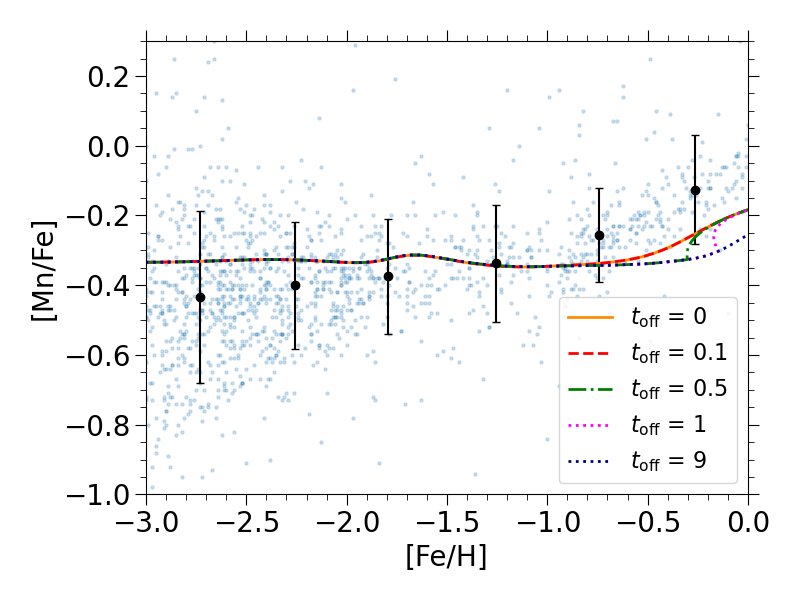}
    \caption{[Mn/Fe] versus [Fe/H] for the best-fit model L25-SK18-RnoKH with \( D_{\mathrm{PBH}} \) set to its best-fit value. The curves show the effect of varying the switch-off time (\( t_{\mathrm{off}} \)) of PBH-induced SNe Ia: \( t_{\mathrm{off}} = 0 \) (orange line), \( 0.1\,\mathrm{Gyr} \) (red dashed line), \( 0.5\,\mathrm{Gyr} \) (green dash-dotted line), and \( 1\,\mathrm{Gyr} \) (magenta dotted line). Black points with error bars represent binned observational data from the SAGA database, and small blue points represent star measurements.}
    \label{fig:GCE_time}
\end{figure}

In Figure \ref{fig:GCE_dpbh}, we show the comparative study for the best model (L25-SK18-RnoKH), against different fractions of $D_\mathrm{PBH}$, with the same $f_\mathrm{single}$. We consider two different fractions: one in which no PBH-induced SNe Ia occur ($D_\mathrm{PBH}$ = 0), and one in which only SNe Ia are induced by non-binary systems ($D_\mathrm{PBH}$ = 1). By examining [Mn/Fe] alone, $D_{\rm PBH} = 0$ yields the trend that best fits the data within the error bars across all [Fe/H] values. This can be contrasted with the $D_\mathrm{PBH} = 1$ case (red dashed line), which tends to overproduce [Mn/Fe] in $-2.0 \leq \mathrm{[Fe/H]} \leq -1.5$. The best-fit model with an intermediate $D_\mathrm{PBH}$ fraction (orange solid line) shares a similar trend with the $D_{\rm PBH} = 0$ case but deviates at [Fe/H]$\sim -1$, and it rises more slowly than the observed trend. 

We remark that the best-fit model is determined by fitting all elements considered, including [Mn/Fe] and [Ni/Fe]. In the bottom panel, we plot [Ni/Fe] against [Fe/H]. In the extreme case $D_{\rm PBH} = 1$, the rise of the Ni/Fe clearly deviates from the global trend of the data. Meanwhile, for $D_{\rm PBH} = 0$, the curve has a bump near [Fe/H]$\sim -0.5$ compared to the best-fit model. Since the error bars for [Ni/Fe] are generally smaller, this favours the non-zero $D_{\rm PBH}$ over the no PBH case in the $\chi^2$ approach.

In Figure \ref{fig:GCE_fsing}, we show another comparative study for the best model (L25-SK18-RnoKH), against different fractions of $f_\mathrm{single}$, with the same $D_\mathrm{PBH}$. We similarly presume two different fractions, one such that none of the WD systems are non-binary ($f_\mathrm{single}$ = 0), and one where all WD systems are non-binary ($f_\mathrm{single}$ = 1). We can see that for $-3.0 \leq \mathrm{[Fe/H]} \leq -0.75$, [Mn/Fe] production remains about the same for all three lines. This is most likely due to $D_\mathrm{PBH}$ being the same for each. 

In Figure \ref{fig:GCE_time}, we show the comparative study for the best model (L25-SK18-RnoKH), comparing different switch-off times ($t_{\mathrm{off}}$) for the PBH-induced SNe Ia channel. We consider four cases: an immediate switch-off ($t_{\mathrm{off}} = 0$ Gyr), and delayed switch-offs at $t_{\mathrm{off}} = 0.1$ Gyr to 9 Gyr. For $-3.0 \leq \mathrm{[Fe/H]} \leq -1.0$, [Mn/Fe] production remains nearly identical for all four cases, and this indicates that early Galactic Mn enrichment is insensitive to the PBH channel switch-off time. This is because the early massive star evolution produces the baseline metallicity, which determines the regime where PBH matters. This parameter also changes the [Mn/Fe] in recent universe only. Specifically, with $t_{\mathrm{off}}<1$ Gyr producing a smoother rise in [Mn/Fe] toward $\approx -0.2$, while $t_{\mathrm{off}}=9$ Gyr exhibits a suppressed curve until it catches up with other curves at a higher metallicity.

\subsection{Effects of Simmering Phase on Mn Production}

In Table \ref{tab:GCE_model}, we have included models where the WDs do not experience the simmering phase. This is possible for the PBH channel because the WD could still ignite without interacting with the binary companion. Without the simmering phase (NS models), the abundant $^{14}$N created in the CNO cycle cannot be converted into $^{22}$Ne, which suppresses the later production of $^{55}$Mn and $^{58}$Ni. Indeed, in the table we observe a lower $\chi^2$ value for the NS models compared to the equivalent model with simmering. 

\begin{figure}
    \centering
    \includegraphics[width=0.95\linewidth]{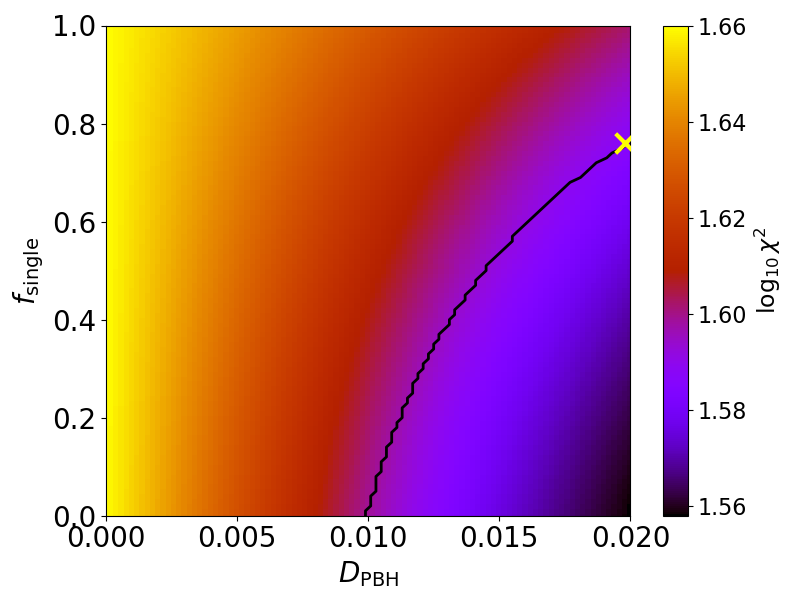}
    \caption{This color plot shows the best no-simmering model: L25-SK18-BnoKH-NS. Each point represents a different model run with a different parameter set of $D_\mathrm{PBH}$ and $f_{\rm single}$. The yellow cross marks the parameters for the overall best-fit model.}
    \label{fig:GCE_chisq_ns}
\end{figure}

In Figure \ref{fig:GCE_chisq_ns} we plot the $\chi^2$ colour plot similar to Figure \ref{fig:GCE_chisq} but for L25-SK18-BnoKH-NS. The suppressed stable Mn and Ni production results in a more relaxed parameter space up to a higher value of $D_{\rm PBH}$. 

\begin{figure}
    \centering
    \includegraphics[width=0.95\linewidth]{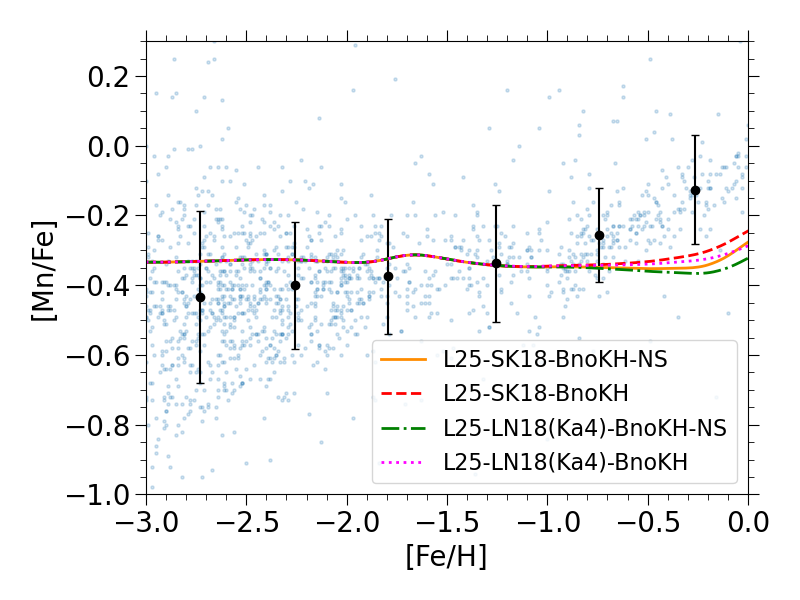}
    \caption{[Mn/Fe] versus [Fe/H] for the BnoKH models, L25-LN18(Ka4)-BnoKH (orange line), L25-SK18-BnoKH (red dashed line), and their no simmering counterparts, (L25-LN18(Ka4)-BnoKH-NS (green dash-dotted line), and L25-SK18-BnoKH-NS (magenta dotted line). Black points with error bars represent binned observational data from the SAGA database, and small blue points represent star measurements.}
    \label{fig:GCE_nosim}
\end{figure}

In Figure \ref{fig:GCE_nosim} we also plot the evolution of [Mn/Fe] of the NS models, and their counterpart models where simmering is assumed. The rise of the Mn near $Z_{\odot}$ for the best models occurs at a similar rate, but the rise occurs later for the NS models, near [Fe/H] $\sim 0.0$. Despite the difference, the rise near $Z_{\odot}$ remains insufficient because the SN Ia channel is inherently suppressed based on the supersolar Ni abundance. 

\section{Discussion}
\label{sec:discussion}

\subsection{Comparing Supernova Rates}

The PBH channel introduces a new dimension to the SN Ia family, especially to early supernovae. We define
\begin{equation}
    f_{\rm PBH} = \frac{r_{\rm PBH}}{r_{\rm PBH} + r_{\rm bin}}, f_{\rm Ia} = \frac{r_{\rm PBH} + r_{\rm bin}}{r_{\rm PBH} + r_{\rm bin} + r_{\rm CCSN}}, 
\end{equation}
to characterize the relative importance of the PBH and the SN Ia channels. Here $r_{\rm PBH}$, $r_{\rm bin}$ and $r_{\rm CCSN}$ are the rates of PBH-induced SNe Ia, SNe Ia from binary systems and massive star explosions obtained from the GCE models, in units of events/yr/Mpc$^{3}$.

\begin{figure}
    \centering
    \includegraphics[width=0.95\linewidth]{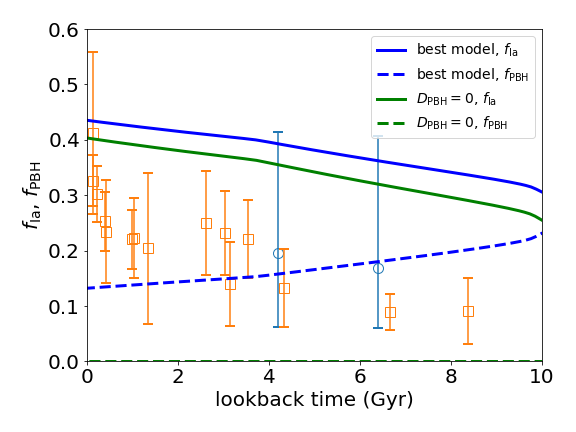}
    \caption{The time evolution of $f_{\rm Ia}$ (blue solid line) and $f_{\rm PBH}$ (blue dashed line) for the model L25-SK18-BnoKH and for comparative model with $D_{\rm PBH} = 0$ set by hand (green solid line). Note that in this model we do not include $t_{\rm off}$ to avoid the sudden drop of the rate. The circles are the $f_{\rm Ia}$ derived from supernova surveys derived from \cite{Melinder2012SNRateZ}, and squares derived from \cite{Maoz2017DTD, Pessi2025CCSNRate}.}
    \label{fig:rate_ratios}
\end{figure}

In Figure \ref{fig:rate_ratios}, we compare ratios of supernova ratio $f_{\rm Ia}$ to show the evolution of different types of SNe over the history of the universe for the best-fit model in L25-SK18-BnoKH. To demonstrate the evolution of the PBH channel of the entire cosmic history, here we exclude $t_{\rm off}$ so that the rate evolution is smooth across cosmic time.

The rising $f_{\rm PBH}$ at the early time suggests that most early SNe Ia come from the PBH channel. The binary channel occurs in a lower mass system, which will take a longer time to occur. The fraction for the PBH channel continues to decrease, which agrees with our expectation that the binary system takes time to become the dominant SN Ia channel. We also include the derived $r_{\rm Ia}$ based on the SN survey Stockholm VIMOS Supernova Survey (SVISS) analyzed in \cite{Melinder2012SNRateZ}. In that work, SNe Ia and CCSNe with redshift $0.1-1.0$ are classified. This provides a window for constraining individual SN rates in the early universe. 

We also compare with later analysis performed by \cite{Maoz2017DTD, Pessi2025CCSNRate}, documenting the SN rate as functions of redshifts based on All-Sky Automated Survey for SuperNovae \citep[ASAS-SN II, ][]{Holoien2017ASASSN}, Cluster Lensing And Supernova survey with Hubble \citep[CLASH, ][]{Graur2014CLASH} and Supernova Legacy Survey \citep[SNLS, ][]{Perrett2012SRS}. These data show a larger deviation from our predicted rates. Data up to 4 Gyr from the current universe has the $f_{\rm Ia}$ about two $\sigma$ above the observed data. Beyond that, the $f_{\rm Ia}$ drops significantly to about $\sim 0.1$. Our GCE model predicts a too high rate and a too flat slope in comparison with those data points ($z > 0.4$). This could be directly related to the DTD which we will further elaborate in Section \ref{sec:caveats}.

Despite the direct explosion possibility in the PBH channel, the entire SN Ia family is less important at early times -- agreeing with the standard GCE picture. The SN Ia rate increases steadily to about 40\% of the entire supernova population. Future SN surveys, especially those tracing early SNe, e.g., JWST, could lead to more precise measurements of this ratio and stronger constraints on the parameter $D_{\rm PBH}$.

\subsection{Caveats}
\label{sec:caveats}
In this work, we have developed the metallicity-dependent PBH-triggered SN Ia models. The models are compared with observed SNe Ia, SNR, and GCE. 

The GCE code used in this work does not assume the delay time distribution (DTD) \citep[for those who included, e.g., ][]{Matteucci2006GCESNIa, Matteucci2009GCESNIa} and the metallicity dependence \citep[for those who included, e.g., ][]{Kobayashi2020GCE}. In Equation \ref{eq:GCE}, only the delay-time due to the main-sequence lifetime is included, but not the delay due to binary interaction before the explosion is triggered. This is coupled to our assumption that the SN Ia model contains only one metallicity, which could overestimate the metallicity of the exploding WDs at early times, where the exploding stars originate from a lower metallicity environment. A more consistent and accurate estimation of the PBH channel will require the DTD for the binary channel with a range of metallicity. The DTD could be zero for a single star if they are directly hit by PBH, or about $10^9$ years in single degenerate scenario, or up to a few of $10^9$ years in double degenerate scenario \citep{Greggio2005DTD}.

A direct impact by the delayed time distribution (DTD) is the transition metallicity of [$\alpha$/Fe] where the SNe Ia take over the Fe production \citep{Matteucci2009GCESNIa, Andres2017DTD, Poulhazan2018DTD, Palicio2023DTD, Dubay2024DTD}. The volumetric SN formation history indicates a DTD $\sim t^{-1.1}$ \citep{Maoz2017DTD}. The presence of DTD, coupled with the starburst star formation history, is shown to be the key for explaining the high [$\alpha$/Fe] trends in Milky Way stars. Coupling the DTD, which focuses on the time-domain of the GCE equations, will be an interesting future work to expand the current code, primarily done in the mass-domain.

Another impact from the DTD is the SN rate. In Figure \ref{fig:rate_ratios}, we notice that the slope of the $r_{\rm Ia}$ is much steeper closer to the current universe. While our current explosion model shows flatter curves. A DTD in the order of $\sim3-10$ Gyr is important to suppress early SN Ia rate, and delay their explosion till the current universe. The suppressed SN Ia rate implies that about half of the cosmic Fe in the current universe comes from CCSNe \citep{Maoz2017DTD}. For future fitting, the chemical abundances and the supernova rates should be simultaneous fitting parameters for constraining the supernova models, in comparison with galactic stars, instead of being a posterior parameter.

Another shortcoming in the computation is that we consider the population based on the expectation of forming 0.6 -- 1.1 $M_{\odot}$ WD in the individual star limit. In fact, some binary systems can also carry out mass transfer until they reach the required mass. In the current model, the fraction of SNe Ia in the binary channel assumes both single-star and binary star systems. However, in the PBH channel, the WD doesn't need to accumulate mass all the way up to the Ch-mass. It is possible that a subset of binary systems where the primary WD does not have the required mass, but can later reach the mass by binary interaction. The actual fraction and the rate depend dynamically on the local PBH and WD densities. While our approach can avoid the uncertainty, it will underestimate the actual population of WDs capable of carrying out PBH-triggered SN Ia. A more consistent calculation will require integrating the fraction of binary systems that are capable of accreting mass and reaching the prescribed mass range. This step also determines the $f_{\rm single}$ as a function of cosmic time. 

In this work, we have parametrized the PBH by the rate $D_{\rm PBH}$, its single star fraction, and the switch-off time. The switch-off time spontaneously makes the PBH channel rate zero beyond that time. This simplified model provides a trackable parameter to control its relative importance at different cosmic ages. However, the abrupt transition could be unrealistic. A more consistent approach will take into account the dark matter evolution around the galactic bulge as a function of time and the local WD density. This will allow the $D_{\rm PBH}$ to smoothly transit to zero over time. However, this will also assume a certain DM halo model and the actual motion of the WDs. Such improvement will require the DM profile from cosmological N-body simulation to track the typical DM mass density, especially near the galactic bulge, for obtaining an upper limit for the PBH channel.

\subsection{Conclusion}

In this article, we study the metallicity dependence of the Type Ia supernova models (SNe Ia) triggered by the primordial black hole (PBH) as dark matter. We show that these PBH-triggered SNe Ia behave similarly to ordinary SNe Ia, where higher metallicity flavours the production of Mn and Ni. Such dependence can be used to explain the individual measurement of SN light curves (e.g., SN 2011fe, SN 2012cg) and SN remnants (e.g., Tycho, 3C 397), which measured the $^{57}$Ni/$^{56}$Ni ratio and the Ni/Mn ratios. We show that some of these observed SNe Ia could be the consequence of the PBH-triggered explosion.

We also apply our models in the context of galactic chemical evolution. We include the PBH as a new stellar chemical source in the code. We perform parameter surveys and look for the best parameter sets in the PBH channel that can reproduce the trend of chemical elements for stars in the Milky Way. We show that the SN Ia models in the canonical binary-evolution channel affect the interpreted fraction of PBH-triggered SNe Ia. Despite that, they point at a non-zero fraction of the PBH channel and a non-zero switch-off time. We further studied how the model parameters of this channel, including its fraction, the switch-off time, and its simmering effect, affect the overall chemical trend, especially Mn evolution. For each massive star and binary-channel SN Ia model, the optimal models show little difference from each other. 

Our models suggest that the new channel will not significantly affect the evolution of chemical elements in the early universe. The PBH channel only affects the elemental trends for stars near the $Z_{\odot}$. The channel will be important for galactic sites that exhibit chemical element trends different from those of the Milky Way Galaxy. For example, a falling [Mn/Fe] due to metallicity could be a result of abundant explosions of low-mass SN Ia by PBH. The PBH channel thus could be an alternative to explain the widespread elemental abundances in high-metallicity stars from various stellar surveys. However, some systematic mismatches, such as the low value of [S/Fe] and high value of [Ni/Fe], will require the simultaneous improvement in the massive star explosion models to reconcile with the data pattern. 

\section*{Acknowledgment}

We thank Frank Timmes for the open-source subroutines of the \texttt{Helmholtz} equation of state, the \texttt{torch} nuclear reaction network, and the galactic chemical evolution code. 
This material is based upon work supported by the National Science Foundation under Grant AST-2316807.
K.N. acknowledges support by the World Premier International Research Center Initiative (WPI), and JSPS KAKENHI Grant Numbers JP20K04024, JP21H04499, JP23K03452, and JP25K01046. 

\software{  Numpy \citep{Numpy},
            Matplotlib \citep{Matplotlib},
            Pandas \citep{Pandas}
          }

\vspace{5mm}

\appendix

\section{Yield Table of SN Ia triggered by PBH}
\label{sec:appA}
In the main text, we have presented the isotopic and elemental distribution of the SNe Ia triggered by PBH. Here we provide the detailed nucleosynthetic yield table, which could be applicable in direct comparison with other observational data such as SNRs, late-time LCs, and spectra. Here we provide the table form of the supernova ejecta masses for the characteristic model 06B-noKH with different initial metallicity in Table \ref{table:yield_isotopes} for the stable isotopes, Table \ref{table:unstable_isotopes} for the unstable short-lived isotopes, and Table \ref{table:yield_elements} for the elements \ref{table:yield_elements}. 

\begin{longtable}[h]{c c c c c c c}
 \caption{The isotope yields table for the characteristic benchmark model 06B-noKH at various metallicity. Masses are in units of $M_{\odot}$.\label{table:yield_isotopes}} \\
 
 Isotope &  $Z = 0$ & $Z = 0.002$ & $Z = 0.01$ & $Z = 0.02$ & $Z = 0.06$ & $Z = 0.10$\\
 \hline
 \endfirsthead

 \multicolumn{7}{c}{\textit{Continuation of Table \ref{table:yield_isotopes}.}} \\
 Isotope & $Z = 0$ & $Z = 0.002$ & $Z = 0.01$ & $Z = 0.02$ & $Z = 0.06$ & $Z = 0.10$ \\
 \hline
 \endhead

 \hline
 \endfoot

 \hline
 \endlastfoot

 $^{12}$C & $2.17 \times 10^{-3}$ & $2.16 \times 10^{-3}$ & $2.14 \times 10^{-3}$ & $2.12 \times 10^{-3}$ & $2.8 \times 10^{-3}$ & $2.5 \times 10^{-3}$ \\
 $^{13}$C & $2.34 \times 10^{-10}$ & $1.21 \times 10^{-10}$ & $6.46 \times 10^{-10}$ & $2.33 \times 10^{-9}$ & $2.74 \times 10^{-8}$ & $8.35 \times 10^{-8}$ \\
 $^{14}$N & $1.2 \times 10^{-8}$ & $4.45 \times 10^{-9}$ & $1.36 \times 10^{-8}$ & $3.5 \times 10^{-8}$ & $3.9 \times 10^{-7}$ & $7.48 \times 10^{-7}$ \\
 $^{15}$N & $2.35 \times 10^{-6}$ & $6.17 \times 10^{-8}$ & $3.57 \times 10^{-9}$ & $1.20 \times 10^{-9}$ & $1.12 \times 10^{-9}$ & $4.50 \times 10^{-9}$ \\
 $^{16}$O & $1.28 \times 10^{-1}$ & $1.30 \times 10^{-1}$ & $1.33 \times 10^{-1}$ & $1.35 \times 10^{-1}$ & $1.37 \times 10^{-1}$ & $1.35 \times 10^{-1}$ \\
 $^{17}$O & $1.89 \times 10^{-10}$ & $5.48 \times 10^{-10}$ & $5.9 \times 10^{-9}$ & $1.55 \times 10^{-8}$ & $5.77 \times 10^{-8}$ & $7.54 \times 10^{-8}$ \\
 $^{18}$O & $7.28 \times 10^{-12}$ & $1.64 \times 10^{-11}$ & $9.56 \times 10^{-11}$ & $1.89 \times 10^{-10}$ & $1.65 \times 10^{-9}$ & $6.97 \times 10^{-9}$ \\
 $^{19}$F & $2.61 \times 10^{-10}$ & $4.79 \times 10^{-11}$ & $1.63 \times 10^{-11}$ & $4.15 \times 10^{-11}$ & $3.11 \times 10^{-10}$ & $6.78 \times 10^{-10}$ \\
 $^{20}$Ne & $7.10 \times 10^{-3}$ & $7.6 \times 10^{-3}$ & $6.99 \times 10^{-3}$ & $6.88 \times 10^{-3}$ & $6.37 \times 10^{-3}$ & $5.98 \times 10^{-3}$ \\
 $^{21}$Ne & $1.20 \times 10^{-8}$ & $1.52 \times 10^{-8}$ & $9.99 \times 10^{-8}$ & $3.27 \times 10^{-7}$ & $2.66 \times 10^{-6}$ & $8.24 \times 10^{-6}$ \\
 $^{22}$Ne & $1.38 \times 10^{-8}$ & $1.27 \times 10^{-6}$ & $6.35 \times 10^{-6}$ & $1.26 \times 10^{-5}$ & $4.9 \times 10^{-5}$ & $8.34 \times 10^{-5}$ \\
 $^{23}$Na & $1.16 \times 10^{-5}$ & $1.26 \times 10^{-5}$ & $1.72 \times 10^{-5}$ & $2.50 \times 10^{-5}$ & $6.2 \times 10^{-5}$ & $1.21 \times 10^{-4}$ \\
 $^{24}$Mg & $2.39 \times 10^{-2}$ & $2.24 \times 10^{-2}$ & $1.66 \times 10^{-2}$ & $1.25 \times 10^{-2}$ & $6.27 \times 10^{-3}$ & $4.36 \times 10^{-3}$ \\
 $^{25}$Mg & $3.50 \times 10^{-7}$ & $4.40 \times 10^{-6}$ & $2.53 \times 10^{-5}$ & $6.49 \times 10^{-5}$ & $2.46 \times 10^{-4}$ & $5.4 \times 10^{-4}$ \\
 $^{26}$Mg & $3.79 \times 10^{-6}$ & $1.20 \times 10^{-5}$ & $4.72 \times 10^{-5}$ & $9.7 \times 10^{-5}$ & $3.53 \times 10^{-4}$ & $1.5 \times 10^{-3}$ \\
 $^{26}$Al & $2.80 \times 10^{-29}$ & $2.80 \times 10^{-29}$ & $2.80 \times 10^{-29}$ & $2.80 \times 10^{-29}$ & $9.18 \times 10^{-28}$ & $2.51 \times 10^{-10}$ \\
 $^{27}$Al & $1.32 \times 10^{-4}$ & $3.36 \times 10^{-4}$ & $8.41 \times 10^{-4}$ & $1.4 \times 10^{-3}$ & $1.10 \times 10^{-3}$ & $1.16 \times 10^{-3}$ \\
 $^{28}$Si & $1.52 \times 10^{-1}$ & $1.58 \times 10^{-1}$ & $1.60 \times 10^{-1}$ & $1.59 \times 10^{-1}$ & $1.52 \times 10^{-1}$ & $1.41 \times 10^{-1}$ \\
 $^{29}$Si & $7.12 \times 10^{-5}$ & $3.21 \times 10^{-4}$ & $6.49 \times 10^{-4}$ & $1.12 \times 10^{-3}$ & $3.49 \times 10^{-3}$ & $6.51 \times 10^{-3}$ \\
 $^{30}$Si & $4.61 \times 10^{-5}$ & $1.0 \times 10^{-4}$ & $9.51 \times 10^{-4}$ & $2.17 \times 10^{-3}$ & $7.56 \times 10^{-3}$ & $1.46 \times 10^{-2}$ \\
 $^{31}$P & $1.20 \times 10^{-4}$ & $7.50 \times 10^{-5}$ & $2.71 \times 10^{-4}$ & $4.68 \times 10^{-4}$ & $1.6 \times 10^{-3}$ & $1.46 \times 10^{-3}$ \\
 $^{32}$S & $7.77 \times 10^{-2}$ & $7.64 \times 10^{-2}$ & $7.76 \times 10^{-2}$ & $7.71 \times 10^{-2}$ & $6.71 \times 10^{-2}$ & $5.58 \times 10^{-2}$ \\
 $^{33}$S & $5.27 \times 10^{-5}$ & $1.37 \times 10^{-4}$ & $2.80 \times 10^{-4}$ & $3.83 \times 10^{-4}$ & $5.65 \times 10^{-4}$ & $5.91 \times 10^{-4}$ \\
 $^{34}$S & $4.66 \times 10^{-6}$ & $1.30 \times 10^{-4}$ & $1.11 \times 10^{-3}$ & $2.48 \times 10^{-3}$ & $8.45 \times 10^{-3}$ & $1.34 \times 10^{-2}$ \\
 $^{36}$S & $5.46 \times 10^{-12}$ & $1.58 \times 10^{-9}$ & $4.21 \times 10^{-8}$ & $2.48 \times 10^{-7}$ & $9.82 \times 10^{-6}$ & $3.3 \times 10^{-5}$ \\
 $^{35}$Cl & $8.51 \times 10^{-6}$ & $1.95 \times 10^{-5}$ & $8.74 \times 10^{-5}$ & $1.74 \times 10^{-4}$ & $3.22 \times 10^{-4}$ & $2.37 \times 10^{-4}$ \\
 $^{37}$Cl & $3.55 \times 10^{-6}$ & $9.87 \times 10^{-6}$ & $1.81 \times 10^{-5}$ & $2.49 \times 10^{-5}$ & $4.1 \times 10^{-5}$ & $4.36 \times 10^{-5}$ \\
 $^{36}$Ar & $1.63 \times 10^{-2}$ & $1.51 \times 10^{-2}$ & $1.43 \times 10^{-2}$ & $1.33 \times 10^{-2}$ & $1.0 \times 10^{-2}$ & $7.80 \times 10^{-3}$ \\
 $^{38}$Ar & $4.36 \times 10^{-7}$ & $6.82 \times 10^{-5}$ & $5.49 \times 10^{-4}$ & $1.21 \times 10^{-3}$ & $3.73 \times 10^{-3}$ & $5.22 \times 10^{-3}$ \\
 $^{40}$Ar & $4.0 \times 10^{-15}$ & $8.1 \times 10^{-12}$ & $4.2 \times 10^{-10}$ & $4.32 \times 10^{-9}$ & $2.4 \times 10^{-7}$ & $5.43 \times 10^{-7}$ \\
 $^{39}$K & $1.49 \times 10^{-6}$ & $1.50 \times 10^{-5}$ & $5.27 \times 10^{-5}$ & $8.38 \times 10^{-5}$ & $1.24 \times 10^{-4}$ & $1.1 \times 10^{-4}$ \\
 $^{40}$K & $3.81 \times 10^{-12}$ & $1.47 \times 10^{-9}$ & $1.74 \times 10^{-8}$ & $7.88 \times 10^{-8}$ & $2.64 \times 10^{-7}$ & $1.10 \times 10^{-7}$ \\
 $^{41}$K & $4.31 \times 10^{-7}$ & $2.28 \times 10^{-6}$ & $3.86 \times 10^{-6}$ & $5.12 \times 10^{-6}$ & $6.10 \times 10^{-6}$ & $4.71 \times 10^{-6}$ \\
 $^{40}$Ca & $1.62 \times 10^{-2}$ & $1.44 \times 10^{-2}$ & $1.28 \times 10^{-2}$ & $1.16 \times 10^{-2}$ & $8.69 \times 10^{-3}$ & $6.90 \times 10^{-3}$ \\
 $^{42}$Ca & $3.58 \times 10^{-9}$ & $1.66 \times 10^{-6}$ & $1.57 \times 10^{-5}$ & $3.59 \times 10^{-5}$ & $9.88 \times 10^{-5}$ & $1.13 \times 10^{-4}$ \\
 $^{43}$Ca & $7.8 \times 10^{-8}$ & $1.17 \times 10^{-7}$ & $2.67 \times 10^{-7}$ & $2.36 \times 10^{-7}$ & $2.70 \times 10^{-7}$ & $2.77 \times 10^{-7}$ \\
 $^{44}$Ca & $1.39 \times 10^{-5}$ & $1.35 \times 10^{-5}$ & $1.14 \times 10^{-5}$ & $1.0 \times 10^{-5}$ & $7.41 \times 10^{-6}$ & $5.94 \times 10^{-6}$ \\
 $^{46}$Ca & $2.3 \times 10^{-20}$ & $5.42 \times 10^{-15}$ & $3.58 \times 10^{-12}$ & $9.4 \times 10^{-11}$ & $2.58 \times 10^{-9}$ & $2.92 \times 10^{-9}$ \\
 $^{48}$Ca & $2.14 \times 10^{-25}$ & $8.61 \times 10^{-22}$ & $5.87 \times 10^{-18}$ & $8.55 \times 10^{-16}$ & $7.69 \times 10^{-13}$ & $4.5 \times 10^{-12}$ \\
 $^{45}$Sc & $2.89 \times 10^{-8}$ & $7.73 \times 10^{-8}$ & $1.38 \times 10^{-7}$ & $1.95 \times 10^{-7}$ & $2.38 \times 10^{-7}$ & $2.9 \times 10^{-7}$ \\
 $^{46}$Ti & $2.42 \times 10^{-7}$ & $9.21 \times 10^{-7}$ & $7.19 \times 10^{-6}$ & $1.54 \times 10^{-5}$ & $3.56 \times 10^{-5}$ & $3.57 \times 10^{-5}$ \\
 $^{47}$Ti & $2.21 \times 10^{-7}$ & $3.13 \times 10^{-7}$ & $7.19 \times 10^{-7}$ & $9.30 \times 10^{-7}$ & $1.52 \times 10^{-6}$ & $1.92 \times 10^{-6}$ \\
 $^{48}$Ti & $3.9 \times 10^{-4}$ & $2.77 \times 10^{-4}$ & $2.35 \times 10^{-4}$ & $2.7 \times 10^{-4}$ & $1.50 \times 10^{-4}$ & $1.19 \times 10^{-4}$ \\
 $^{49}$Ti & $2.77 \times 10^{-6}$ & $7.19 \times 10^{-6}$ & $1.11 \times 10^{-5}$ & $1.37 \times 10^{-5}$ & $1.74 \times 10^{-5}$ & $1.90 \times 10^{-5}$ \\
 $^{50}$Ti & $7.0 \times 10^{-19}$ & $2.7 \times 10^{-13}$ & $1.2 \times 10^{-10}$ & $7.56 \times 10^{-10}$ & $2.75 \times 10^{-9}$ & $6.52 \times 10^{-9}$ \\
 $^{50}$V & $3.31 \times 10^{-15}$ & $1.7 \times 10^{-11}$ & $1.2 \times 10^{-9}$ & $3.68 \times 10^{-9}$ & $1.4 \times 10^{-8}$ & $3.11 \times 10^{-8}$ \\
 $^{51}$V & $4.40 \times 10^{-6}$ & $9.89 \times 10^{-6}$ & $2.55 \times 10^{-5}$ & $3.66 \times 10^{-5}$ & $6.51 \times 10^{-5}$ & $9.13 \times 10^{-5}$ \\
 $^{50}$Cr & $1.45 \times 10^{-6}$ & $1.11 \times 10^{-5}$ & $6.64 \times 10^{-5}$ & $1.44 \times 10^{-4}$ & $4.49 \times 10^{-4}$ & $6.45 \times 10^{-4}$ \\
 $^{52}$Cr & $4.72 \times 10^{-3}$ & $4.43 \times 10^{-3}$ & $3.85 \times 10^{-3}$ & $3.46 \times 10^{-3}$ & $3.6 \times 10^{-3}$ & $3.70 \times 10^{-3}$ \\
 $^{53}$Cr & $6.67 \times 10^{-5}$ & $1.29 \times 10^{-4}$ & $2.10 \times 10^{-4}$ & $2.67 \times 10^{-4}$ & $4.54 \times 10^{-4}$ & $6.64 \times 10^{-4}$ \\
 $^{54}$Cr & $1.21 \times 10^{-13}$ & $2.25 \times 10^{-10}$ & $6.90 \times 10^{-9}$ & $3.0 \times 10^{-8}$ & $6.7 \times 10^{-7}$ & $2.13 \times 10^{-6}$ \\
 $^{55}$Mn & $7.82 \times 10^{-5}$ & $4.21 \times 10^{-4}$ & $1.6 \times 10^{-3}$ & $1.53 \times 10^{-3}$ & $3.60 \times 10^{-3}$ & $7.84 \times 10^{-3}$ \\
 $^{54}$Fe & $5.11 \times 10^{-5}$ & $9.19 \times 10^{-4}$ & $5.4 \times 10^{-3}$ & $1.2 \times 10^{-2}$ & $3.5 \times 10^{-2}$ & $6.11 \times 10^{-2}$ \\
 $^{56}$Fe & $6.26 \times 10^{-1}$ & $6.21 \times 10^{-1}$ & $6.3 \times 10^{-1}$ & $5.84 \times 10^{-1}$ & $5.16 \times 10^{-1}$ & $4.51 \times 10^{-1}$ \\
 $^{57}$Fe & $8.69 \times 10^{-3}$ & $9.58 \times 10^{-3}$ & $1.21 \times 10^{-2}$ & $1.49 \times 10^{-2}$ & $2.20 \times 10^{-2}$ & $2.47 \times 10^{-2}$ \\
 $^{58}$Fe & $3.49 \times 10^{-13}$ & $2.54 \times 10^{-10}$ & $2.29 \times 10^{-9}$ & $8.38 \times 10^{-9}$ & $9.14 \times 10^{-8}$ & $2.38 \times 10^{-7}$ \\
 $^{60}$Fe & $2.66 \times 10^{-26}$ & $6.33 \times 10^{-20}$ & $5.19 \times 10^{-19}$ & $1.4 \times 10^{-18}$ & $1.25 \times 10^{-18}$ & $7.3 \times 10^{-18}$ \\
 $^{59}$Co & $3.54 \times 10^{-5}$ & $1.4 \times 10^{-4}$ & $3.71 \times 10^{-4}$ & $4.99 \times 10^{-4}$ & $5.11 \times 10^{-4}$ & $3.32 \times 10^{-4}$ \\
 $^{58}$Ni & $4.14 \times 10^{-4}$ & $1.22 \times 10^{-3}$ & $1.10 \times 10^{-2}$ & $2.42 \times 10^{-2}$ & $7.79 \times 10^{-2}$ & $1.18 \times 10^{-1}$ \\
 $^{60}$Ni & $6.84 \times 10^{-3}$ & $7.6 \times 10^{-3}$ & $6.1 \times 10^{-3}$ & $5.1 \times 10^{-3}$ & $2.13 \times 10^{-3}$ & $9.31 \times 10^{-4}$ \\
 $^{61}$Ni & $1.37 \times 10^{-4}$ & $1.49 \times 10^{-4}$ & $1.70 \times 10^{-4}$ & $1.82 \times 10^{-4}$ & $1.42 \times 10^{-4}$ & $8.56 \times 10^{-5}$ \\
 $^{62}$Ni & $4.85 \times 10^{-5}$ & $1.99 \times 10^{-4}$ & $8.30 \times 10^{-4}$ & $1.42 \times 10^{-3}$ & $2.16 \times 10^{-3}$ & $1.78 \times 10^{-3}$ \\
 $^{64}$Ni & $2.14 \times 10^{-19}$ & $9.35 \times 10^{-12}$ & $3.72 \times 10^{-11}$ & $1.49 \times 10^{-14}$ & $6.47 \times 10^{-15}$ & $2.46 \times 10^{-14}$ \\
 $^{63}$Cu & $6.86 \times 10^{-7}$ & $1.48 \times 10^{-6}$ & $7.0 \times 10^{-7}$ & $1.1 \times 10^{-6}$ & $1.72 \times 10^{-6}$ & $1.75 \times 10^{-6}$ \\
 $^{65}$Cu & $6.45 \times 10^{-7}$ & $6.91 \times 10^{-7}$ & $7.83 \times 10^{-7}$ & $8.25 \times 10^{-7}$ & $6.8 \times 10^{-7}$ & $3.94 \times 10^{-7}$ \\
 $^{64}$Zn & $9.76 \times 10^{-5}$ & $6.67 \times 10^{-5}$ & $1.0 \times 10^{-5}$ & $6.93 \times 10^{-6}$ & $2.59 \times 10^{-6}$ & $1.27 \times 10^{-6}$ \\
 $^{66}$Zn & $1.24 \times 10^{-6}$ & $2.82 \times 10^{-6}$ & $8.6 \times 10^{-6}$ & $1.29 \times 10^{-5}$ & $1.83 \times 10^{-5}$ & $1.66 \times 10^{-5}$ \\
 $^{67}$Zn & $8.40 \times 10^{-9}$ & $5.90 \times 10^{-9}$ & $3.98 \times 10^{-9}$ & $8.57 \times 10^{-9}$ & $2.6 \times 10^{-8}$ & $2.48 \times 10^{-8}$ \\
 $^{68}$Zn & $4.28 \times 10^{-7}$ & $1.42 \times 10^{-7}$ & $8.43 \times 10^{-9}$ & $5.37 \times 10^{-9}$ & $7.47 \times 10^{-9}$ & $1.40 \times 10^{-8}$ \\
 $^{70}$Zn & $1.1 \times 10^{-24}$ & $9.12 \times 10^{-16}$ & $5.0 \times 10^{-16}$ & $1.12 \times 10^{-17}$ & $1.42 \times 10^{-20}$ & $2.5 \times 10^{-24}$ \\

\end{longtable}

\begin{longtable}[c]{c c c c c c c}
 \caption{Same as Table \ref{table:yield_isotopes} but for short-lived radioactive isotopes. \label{table:unstable_isotopes}} \\
 
 Isotope & $Z = 0$ & $Z = 0.002$ & $Z = 0.01$ & $Z = 0.02$ & $Z = 0.06$ & $Z = 0.10$ \\
 \hline
 \endfirsthead

 \multicolumn{7}{c}{\textit{Continuation of Table \ref{table:unstable_isotopes}.}} \\
 Isotope & $Z = 0$ & $Z = 0.002$ & $Z = 0.01$ & $Z = 0.02$ & $Z = 0.06$ & $Z = 0.10$  \\
 \hline
 \endhead

 \hline
 \endfoot

 \hline
 \endlastfoot

  $^{22}$Na & $9.96 \times 10^{-9}$ & $1.7 \times 10^{-8}$ & $1.95 \times 10^{-8}$ & $2.14 \times 10^{-8}$ & $1.36 \times 10^{-8}$ & $9.75 \times 10^{-9}$ \\
 $^{26}$Al & $3.9 \times 10^{-6}$ & $6.90 \times 10^{-6}$ & $1.25 \times 10^{-5}$ & $1.25 \times 10^{-5}$ & $5.9 \times 10^{-6}$ & $2.35 \times 10^{-6}$ \\
 $^{39}$Ar & $1.8 \times 10^{-13}$ & $1.19 \times 10^{-10}$ & $2.37 \times 10^{-9}$ & $1.37 \times 10^{-8}$ & $1.62 \times 10^{-7}$ & $1.80 \times 10^{-7}$ \\
 $^{40}$K & $3.83 \times 10^{-12}$ & $1.48 \times 10^{-9}$ & $1.75 \times 10^{-8}$ & $7.92 \times 10^{-8}$ & $2.65 \times 10^{-7}$ & $1.10 \times 10^{-7}$ \\
 $^{41}$Ca & $3.76 \times 10^{-7}$ & $1.97 \times 10^{-6}$ & $3.69 \times 10^{-6}$ & $4.95 \times 10^{-6}$ & $5.97 \times 10^{-6}$ & $4.56 \times 10^{-6}$ \\
 $^{44}$Ti & $1.40 \times 10^{-5}$ & $1.26 \times 10^{-5}$ & $1.9 \times 10^{-5}$ & $9.62 \times 10^{-6}$ & $6.72 \times 10^{-6}$ & $5.17 \times 10^{-6}$ \\
 $^{48}$V & $9.96 \times 10^{-10}$ & $5.18 \times 10^{-9}$ & $1.98 \times 10^{-8}$ & $4.2 \times 10^{-8}$ & $9.92 \times 10^{-8}$ & $9.37 \times 10^{-8}$ \\
 $^{49}$V & $5.42 \times 10^{-11}$ & $2.68 \times 10^{-9}$ & $4.16 \times 10^{-8}$ & $1.5 \times 10^{-7}$ & $5.7 \times 10^{-7}$ & $1.10 \times 10^{-6}$ \\
 $^{53}$Mn & $4.55 \times 10^{-9}$ & $1.3 \times 10^{-7}$ & $1.81 \times 10^{-6}$ & $7.8 \times 10^{-6}$ & $1.0 \times 10^{-4}$ & $2.16 \times 10^{-4}$ \\
 $^{60}$Fe & $7.81 \times 10^{-26}$ & $9.96 \times 10^{-19}$ & $8.45 \times 10^{-18}$ & $1.71 \times 10^{-17}$ & $1.94 \times 10^{-17}$ & $9.87 \times 10^{-17}$ \\
 $^{56}$Co & $1.37 \times 10^{-7}$ & $5.83 \times 10^{-7}$ & $2.0 \times 10^{-6}$ & $4.2 \times 10^{-6}$ & $1.53 \times 10^{-5}$ & $4.77 \times 10^{-5}$ \\
 $^{57}$Co & $8.24 \times 10^{-10}$ & $1.16 \times 10^{-7}$ & $1.18 \times 10^{-6}$ & $3.75 \times 10^{-6}$ & $2.70 \times 10^{-5}$ & $4.69 \times 10^{-5}$ \\
 $^{60}$Co & $6.34 \times 10^{-19}$ & $4.81 \times 10^{-14}$ & $1.39 \times 10^{-13}$ & $6.39 \times 10^{-14}$ & $1.9 \times 10^{-12}$ & $5.27 \times 10^{-12}$ \\
 $^{56}$Ni & $6.26 \times 10^{-1}$ & $6.21 \times 10^{-1}$ & $6.3 \times 10^{-1}$ & $5.84 \times 10^{-1}$ & $5.15 \times 10^{-1}$ & $4.48 \times 10^{-1}$ \\
 $^{57}$Ni & $8.69 \times 10^{-3}$ & $9.59 \times 10^{-3}$ & $1.21 \times 10^{-2}$ & $1.49 \times 10^{-2}$ & $2.20 \times 10^{-2}$ & $2.47 \times 10^{-2}$ \\
 $^{59}$Ni & $5.55 \times 10^{-9}$ & $5.55 \times 10^{-8}$ & $4.17 \times 10^{-7}$ & $1.22 \times 10^{-6}$ & $1.48 \times 10^{-5}$ & $5.29 \times 10^{-5}$ \\
 $^{63}$Ni & $1.40 \times 10^{-21}$ & $2.21 \times 10^{-12}$ & $2.99 \times 10^{-12}$ & $4.29 \times 10^{-15}$ & $1.16 \times 10^{-14}$ & $6.91 \times 10^{-14}$ \\

 \end{longtable}

 \begin{longtable}[c]{c c c c c c c}
 \caption{Same as Table \ref{table:yield_isotopes} but for the elemental yields. \label{table:yield_elements}} \\
 
 Isotope & $Z = 0$ & $Z = 0.002$ & $Z = 0.01$ & $Z = 0.02$ & $Z = 0.06$ & $Z = 0.10$  \\
 \hline
 \endfirsthead

 \multicolumn{7}{c}{\textit{Continuation of Table \ref{table:yield_elements}.}} \\
 Isotope & $Z = 0$ & $Z = 0.002$ & $Z = 0.01$ & $Z = 0.02$ & $Z = 0.06$ & $Z = 0.10$  \\
 \hline
 \endhead

 \hline
 \endfoot

 \hline
 \endlastfoot

 C & $2.17 \times 10^{-3}$ & $2.16 \times 10^{-3}$ & $2.14 \times 10^{-3}$ & $2.12 \times 10^{-3}$ & $2.8 \times 10^{-3}$ & $2.5 \times 10^{-3}$ \\
 N & $2.36 \times 10^{-6}$ & $6.62 \times 10^{-8}$ & $1.72 \times 10^{-8}$ & $3.17 \times 10^{-8}$ & $3.10 \times 10^{-7}$ & $7.52 \times 10^{-7}$ \\
 O & $1.28 \times 10^{-1}$ & $1.30 \times 10^{-1}$ & $1.33 \times 10^{-1}$ & $1.35 \times 10^{-1}$ & $1.37 \times 10^{-1}$ & $1.35 \times 10^{-1}$ \\
 F & $2.61 \times 10^{-10}$ & $4.79 \times 10^{-11}$ & $1.63 \times 10^{-11}$ & $4.15 \times 10^{-11}$ & $3.11 \times 10^{-10}$ & $6.78 \times 10^{-10}$ \\
 $Ne$ & $7.10 \times 10^{-3}$ & $7.6 \times 10^{-3}$ & $6.99 \times 10^{-3}$ & $6.89 \times 10^{-3}$ & $6.41 \times 10^{-3}$ & $6.7 \times 10^{-3}$ \\
 Na & $1.16 \times 10^{-5}$ & $1.26 \times 10^{-5}$ & $1.72 \times 10^{-5}$ & $2.50 \times 10^{-5}$ & $6.2 \times 10^{-5}$ & $1.21 \times 10^{-4}$ \\
 Mg & $2.39 \times 10^{-2}$ & $2.24 \times 10^{-2}$ & $1.67 \times 10^{-2}$ & $1.27 \times 10^{-2}$ & $6.87 \times 10^{-3}$ & $5.92 \times 10^{-3}$ \\
 Al & $1.32 \times 10^{-4}$ & $3.36 \times 10^{-4}$ & $8.41 \times 10^{-4}$ & $1.4 \times 10^{-3}$ & $1.10 \times 10^{-3}$ & $1.16 \times 10^{-3}$ \\
 Si & $1.52 \times 10^{-1}$ & $1.58 \times 10^{-1}$ & $1.62 \times 10^{-1}$ & $1.63 \times 10^{-1}$ & $1.63 \times 10^{-1}$ & $1.63 \times 10^{-1}$ \\
 P & $1.20 \times 10^{-4}$ & $7.50 \times 10^{-5}$ & $2.71 \times 10^{-4}$ & $4.68 \times 10^{-4}$ & $1.6 \times 10^{-3}$ & $1.46 \times 10^{-3}$ \\
 S & $7.78 \times 10^{-2}$ & $7.66 \times 10^{-2}$ & $7.90 \times 10^{-2}$ & $8.0 \times 10^{-2}$ & $7.61 \times 10^{-2}$ & $6.99 \times 10^{-2}$ \\
 Cl & $1.20 \times 10^{-5}$ & $2.94 \times 10^{-5}$ & $1.5 \times 10^{-4}$ & $1.99 \times 10^{-4}$ & $3.62 \times 10^{-4}$ & $2.81 \times 10^{-4}$ \\
 Ar & $1.63 \times 10^{-2}$ & $1.52 \times 10^{-2}$ & $1.48 \times 10^{-2}$ & $1.45 \times 10^{-2}$ & $1.38 \times 10^{-2}$ & $1.30 \times 10^{-2}$ \\
 K & $1.92 \times 10^{-6}$ & $1.73 \times 10^{-5}$ & $5.65 \times 10^{-5}$ & $8.90 \times 10^{-5}$ & $1.30 \times 10^{-4}$ & $1.6 \times 10^{-4}$ \\
 Ca & $1.62 \times 10^{-2}$ & $1.44 \times 10^{-2}$ & $1.28 \times 10^{-2}$ & $1.16 \times 10^{-2}$ & $8.79 \times 10^{-3}$ & $7.2 \times 10^{-3}$ \\
 Sc & $2.89 \times 10^{-8}$ & $7.73 \times 10^{-8}$ & $1.38 \times 10^{-7}$ & $1.95 \times 10^{-7}$ & $2.38 \times 10^{-7}$ & $2.9 \times 10^{-7}$ \\
 Ti & $3.12 \times 10^{-4}$ & $2.85 \times 10^{-4}$ & $2.54 \times 10^{-4}$ & $2.37 \times 10^{-4}$ & $2.5 \times 10^{-4}$ & $1.76 \times 10^{-4}$ \\
 V & $4.40 \times 10^{-6}$ & $9.89 \times 10^{-6}$ & $2.55 \times 10^{-5}$ & $3.66 \times 10^{-5}$ & $6.51 \times 10^{-5}$ & $9.13 \times 10^{-5}$ \\
 Cr & $4.79 \times 10^{-3}$ & $4.57 \times 10^{-3}$ & $4.13 \times 10^{-3}$ & $3.87 \times 10^{-3}$ & $3.97 \times 10^{-3}$ & $5.1 \times 10^{-3}$ \\
 Mn & $7.82 \times 10^{-5}$ & $4.21 \times 10^{-4}$ & $1.6 \times 10^{-3}$ & $1.53 \times 10^{-3}$ & $3.60 \times 10^{-3}$ & $7.84 \times 10^{-3}$ \\
 Fe & $6.35 \times 10^{-1}$ & $6.31 \times 10^{-1}$ & $6.20 \times 10^{-1}$ & $6.9 \times 10^{-1}$ & $5.69 \times 10^{-1}$ & $5.37 \times 10^{-1}$ \\
 Co & $3.54 \times 10^{-5}$ & $1.4 \times 10^{-4}$ & $3.71 \times 10^{-4}$ & $4.99 \times 10^{-4}$ & $5.11 \times 10^{-4}$ & $3.32 \times 10^{-4}$ \\
 Ni & $7.44 \times 10^{-3}$ & $8.63 \times 10^{-3}$ & $1.81 \times 10^{-2}$ & $3.9 \times 10^{-2}$ & $8.23 \times 10^{-2}$ & $1.21 \times 10^{-1}$ \\
 Cu & $1.33 \times 10^{-6}$ & $2.17 \times 10^{-6}$ & $1.48 \times 10^{-6}$ & $1.83 \times 10^{-6}$ & $2.32 \times 10^{-6}$ & $2.14 \times 10^{-6}$ \\
 Zn & $9.93 \times 10^{-5}$ & $6.97 \times 10^{-5}$ & $1.81 \times 10^{-5}$ & $1.98 \times 10^{-5}$ & $2.9 \times 10^{-5}$ & $1.79 \times 10^{-5}$ \\

\end{longtable}

\section{Abundance Graphs}
\label{sec:appB}

\begin{figure}
\centering
\includegraphics[width=0.78\linewidth]{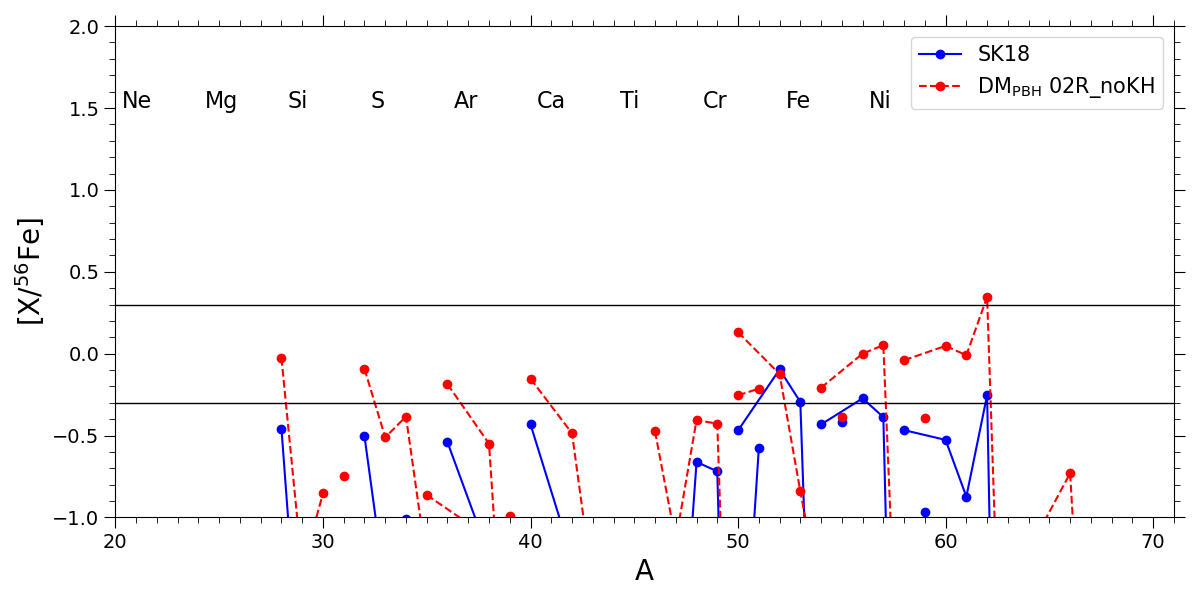}
\includegraphics[width=0.78\linewidth]{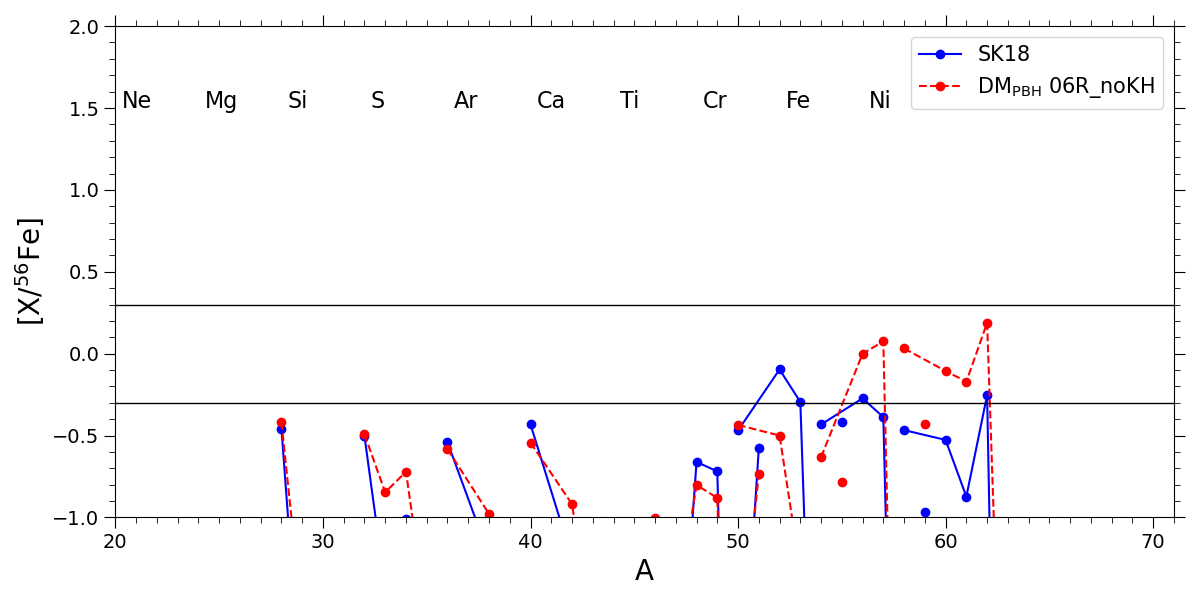}
\includegraphics[width=0.78\linewidth]{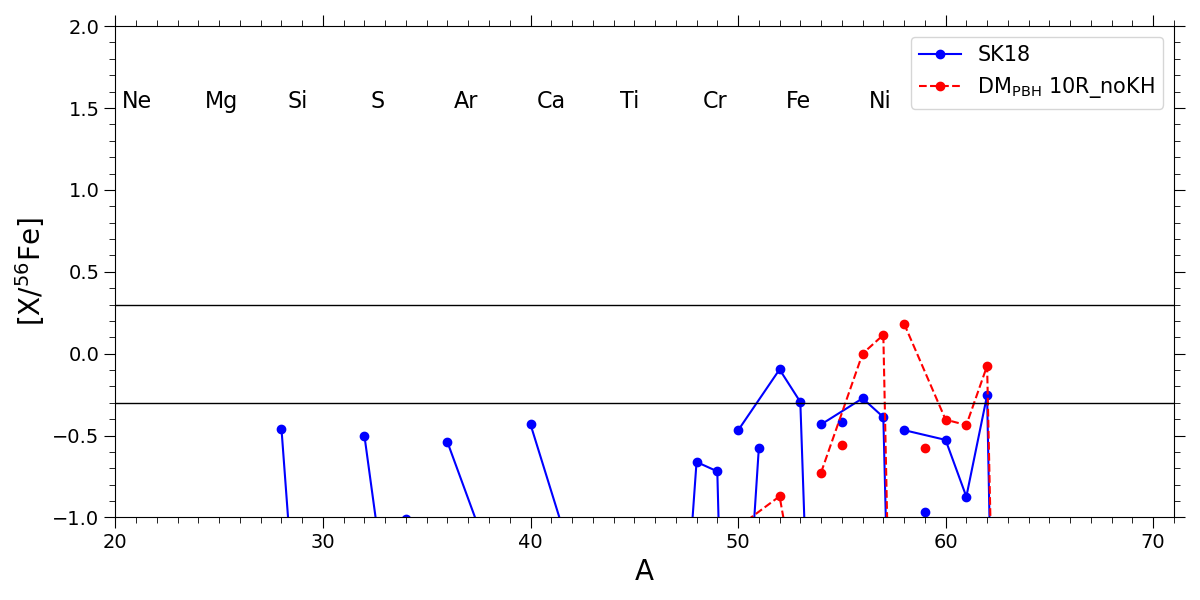}
\caption{(top panel) Isotopic mass fraction ratios [X/$^{56}$Fe] for 02R\_noKH and the subCh-mass SN Ia model SK18.  The horizontal lines indicate twice (upper line) and half (lower line) the solar values. (middle panel) Same as the top panel, but for 06R\_noKH and SK18.  (bottom panel) Same as the top panel, but for 10R\_noKH and SK18.}
\label{fig:abundance_best}
\end{figure}

In Figure \ref{fig:abundance_best}, we show the chemical abundance pattern plots of the PBH-triggered SN Ia models and compared with one of the best SN Ia models for the GCE calculations in Paper II. The mass dependence for the PBH model is clear. Only for low-mass models, where $^{56}$Ni is below normal, the explosions produce both Si-group and Fe-group elements. As mass increases, the Si-group elements become less prominent, leaving the ejecta more dominated by Fe-group elements.


\bibliographystyle{aasjournal}
\pagestyle{plain}
\bibliography{biblio}

\end{document}